%% file: beir9101.tex
\documentclass{aa}
\usepackage{times,graphicx}
\newcommand{\Halpha}{\ensuremath{{\mathrm H}\alpha}}
\newcommand{\oi}{\ion{O}{i}}
\newcommand{\oxir}{\ion{O}{i} 7772$-$5\,{\AA}}
\newcommand{\feii}{\ion{Fe}{ii}}
\newcommand{\Teff}{\ensuremath{T_\mathrm{eff}}}

\newcommand{\Ae}{\ensuremath{\mathrm{AbEm}}}
\newcommand{\Ea}{\ensuremath{\mathrm{EmAb}}}
\newcommand{\Em}{\ensuremath{\mathrm{Em}}}
\newcommand{\Sh}{\ensuremath{\mathrm{Sh}}}
\newcommand{\B}{\ensuremath{\mathrm{B}}}
\newcommand{\Ab}{\ensuremath{\mathrm{Ab}}}
\newcommand{\Be}{\ensuremath{\mathrm{Be}}}

\newcommand{\HEROS}{{\em HEROS}}

\begin{document}

\title{Observations of
{\Halpha},
iron, and oxygen lines in B, Be, and shell stars}
\author{
	S.~M.~Saad\inst{1,2},			
	J.~Kub\'{a}t\inst{1},			
	D.~Kor\v{c}\'{a}kov\'{a}\inst{1},	
	P.~Koubsk\'{y}\inst{1},			
	P.~\v{S}koda\inst{1},			
	M.~\v{S}lechta\inst{1},			
	A.~Kawka\inst{1},			
	A.~Budovi\v{c}ov\'a\inst{1},		
	V.~Votruba\inst{3,1},			
	L.~\v{S}arounov\'a\inst{1},
	M.~I.~Nouh\inst{2}
	}
\authorrunning{S. M. Saad et al.}

\institute{Astronomick\'y \'ustav,
	Akademie v\v{e}d \v{C}esk\'e republiky,
	CZ-251 65 Ond\v{r}ejov, Czech Republic
	\and
	National Research Institute of Astronomy and Geophysics,
	11421 Helwan, Cairo, Egypt
	\and
	\'Ustav teoretick\'e fyziky a astrofyziky P\v{r}F MU,
	Kotl\'a\v{r}sk\'a 2, CZ-611 37 Brno, Czech Republic
	}

\offprints{J. Kub\'at, \\
\email{kubat@sunstel.asu.cas.cz}}

\date{Received 23 August 2004 / Accepted 21 December 2005}

\abstract{We have carried out a spectroscopic survey of several {\B},
{\Be}, and shell stars in optical and near-infrared regions.
Line profiles of the {\Halpha} line and of selected {\feii} and {\oi}
lines are presented.
}

\maketitle

\section{Introduction}

Observations of B, Be, and shell stars in different spectral regions
are important for putting constraints on modeling these stars.
Echelle spectrographs as well as the high sensitivity of 
modern detectors in the red part of the spectrum provide a wealth 
of the information contained in the whole visible
and near infrared regions obtained simultaneously.
The aim of this paper is a spectroscopic survey of line profiles of
{\Halpha} and selected non-hydrogenic lines of iron and oxygen in the
visual and near infrared region for selected bright B and Be stars.
We mainly use the echelle observations secured using the
{\HEROS} ({\em H}eidelberg {\em E}xtended {\em R}ange {\em O}ptical
{\em S}pectrograph) spectrograph attached to the Ond\v{r}ejov 2m
telescope supplemented by several CCD coud\'e spectra.

\section{Observations and data reduction}

The present data are based on new spectroscopic observations of 
13 {\B} stars, 28 {\Be} stars, and 8 shell stars.
Our sample of stars contains objects of spectral types B0 -- B9.5 and
luminosity classes III, IV and V.
Table \ref{hvezdy} summarises the basic information about the observed
objects, their HD and HR no's, their name (if available), MK
spectral type, luminosity class, and Julian dates of the observations.
Some of these stars have never been observed before (to the best of our
knowledge) in the near-infrared region.
The spectra of these stars were obtained between January 2001 and
November 2003 using the fiber-fed echelle spectrograph {\HEROS}
(for a brief description see \v{S}tefl \& Rivinius \cite{SR00}, 
\v{S}koda \& \v{S}lechta \cite{herpopis}) attached to the Cassegrain
focus of the 2m telescope at Ond\v{r}ejov Observatory. 
All the basic data reduction processing, including bias subtraction,
flat fielding, and wavelength calibration, have been done using
the {\HEROS} pipeline written by O. Stahl and A. Kaufer as an extension
of basic MIDAS echelle context (see Stahl et al \cite{hermidas},
also \v{S}koda \& \v{S}lechta \cite{herred}).
Additional observations of several stars were secured using
a CCD detector of a coud\'e spectrograph of the same telescope
(\v{S}lechta \& \v{S}koda \cite{ccdpopis}) and the data were reduced
by the IRAF package.

\input{seznam04}

\section{Observed lines}

\subsection{The hydrogen {\Halpha} line}

The {\Halpha} line in the stars of our sample was found to be either 
in absorption or in emission, however, an intermediate case between
emission and absorption was also found.
Our set of stars exhibits basically five different shapes of {\Halpha} 
line.
We introduce corresponding stellar subclasses for our sample, namely
\renewcommand{\theenumi}{\roman{enumi}}
\begin{enumerate}

\item
Emission is present, but it is completely below the continuum level.
We will denote this subclass as {\Ae} (absorption with emission).

\item
The absorption part is below the level of continuum while
the emission peak is above the continuum.
The ratio of these two parts often varies at different phases for the
same star.
We will denote this subclass as {\Ea} (emission with absorption).

\item
The whole emission feature is completely above the continuum level.
They will be denoted by {\Em} (pure emission).

\item We may also define the shell stars characterised by extremely
sharp {\Halpha} absorption cores as one of the {\Be} star phases. 
Since there is no intrinsic difference between {\Be} and shell stars, 
we denoted them by {\Sh}.
Besides the shape of the {\Halpha} emission line profile, the presence
of other sharp photospheric absorption lines which will be referred to
as ``shell lines'' were also used for defining this subgroup.

\item The last subclass consists of normal {\B} stars with {\Halpha} 
absorption lines.
They are included in our analysis as a kind of standard star.
We denote them by {\Ab}.

\end{enumerate}

Our five subclasses represent a qualitative estimate of the {\Halpha} line
profile in our {\em present} observations.
Table \ref{hvezdy} lists the stars according to these subclasses.
 
\subsection{Non-hydrogen emission lines}

There is a number of non-hydrogen emission lines in the spectra of 
{\Be} stars. The most striking emission features are produced by iron,
helium, and oxygen.
In addition to the hydrogen lines, these non-hydrogenic lines carry
supplementary information coming from different regions of the stellar
envelope depending on their depth of formation.

\subsubsection{Iron lines}

Iron lines bring a lot of information from the circumstellar envelope
depending on their depth of formation.
They are often present in the spectra of Be stars.
An interesting fact about {\feii} lines is that they appear in
the emission spectrum, however, in the approximate theoretical spectrum,
which was calculated under the simplified assumption of a LTE static
plane-parallel atmosphere, they are completely missing.
This indicates that the excitation mechanism is connected more probably
with the atomic structure of {\feii} and the corresponding NLTE pumping than
with some global density changes or even iron overabundance.
However, such a conclusion is to be verified by detailed NLTE
calculations.
We selected several strong lines mostly from the quartet system for our
measurements.

Although the list of available {\feii} lines is quite long, only
several of them are useful for further analysis, in particular the
4233\,{\AA} line from the multiplet (27), the 4584\,{\AA} line from the
multiplet (38), 6148 and 6456\,{\AA} lines from the multiplet (74), and
7516 and 7712\,{\AA} lines from the multiplet (73).
Note that the 6148\,{\AA} line itself is a blend of two neighbouring
{\feii} lines.
Due to its vicinity to the {\Halpha} line the {\feii} 6456\,{\AA} line
was present in the {\Halpha} spectrograms that were obtained with the
CCD coud\'e camera.

The near-infrared line {\feii} 7712\,{\AA} was severely contaminated by
telluric lines in the band 7600 -- 7700\,{\AA}.
However, this line was sometimes strong enough to be measurable in 13
stars of our sample.
In ten of them the line was in emission. 
Other {\feii} lines, namely the 7516, 6456, and 6148\,{\AA} lines are 
also in emission.
However, another {\feii} line, which is not listed in the multiplet
tables of Moore (\cite{multiplets}),
appeared to be quite strong in the synthetic spectra and may be
possibly misidentified with the {\feii} 7516\,{\AA} line.
This line has a wavelength of 7513.162\,{\AA} and arises from the 
transition 
$5s~e\,\element[\mathrm{e}][6][][9/2]{D} \rightarrow
5p~w\,\element[\mathrm{o}][6][][7/2]{P}$.
 For simplicity we will identify this feature with the wavelength
7516\,{\AA} in the figures.
In most cases the iron line {\feii} 6516\,{\AA} was not detectable due
to the huge number of telluric lines.

\input{carym}
\subsubsection{The oxygen infrared triplet \oxir}

The most dominant oxygen line in the visible and near infrared spectra
of {\Be} stars is the near IR triplet line at 7772, 7774, 7775\,{\AA}
emerging from the transition
$3s~\element[\mathrm{o}][5]{S} \rightarrow 3p~\element[][5]{P}$.
Other transitions between quintet levels which fall into the visual
region are missing in the spectra.

\section{Description of the online material}
Results of our observations are plotted in the online Appendix.
It contains an atlas of the individual line profiles sorted and ordered
according to the Table~\ref{hvezdy}.
For each star, the {\Halpha} and {\oxir} lines are plotted.
Iron lines are plotted for each observation where they were available.
The list of {\feii} lines presented is listed in the Table~\ref{cary}.

%

\begin{acknowledgements}
This research has made use of the NASA's Astrophysics Data System
Abstract Service (Kurtz et al. \cite{ADS1}, Eichhorn et al. \cite{ADS2},
Accomazzi et al. \cite{ADS3}, Grant et al. \cite{ADS4}).
Our work was supported by grants of the Grant Agency of the Czech
Republic 205/02/0445 and 102/02/1000.
Astronomical Institute Ond\v{r}ejov is supported by a project
Z10030501.
\end{acknowledgements}


\newcommand{\Alicante}[1]{in The Be Phenomenon in Early Type Stars, IAU
	Coll. 175, M. A. Smith, H. F. Henrichs, \& J. Fabregat eds., ASP
	Conf. Ser. Vol. 214, p. #1}

\Online

%

\appendix




%
%
%

\onecolumn

\section{Atlas of individual line profiles}\label{profily}

Note that the panels which contain more than 
one profile the spectra are arbitrarly shifted by 0.2, 0.4 and 0.6 
below or above the continuum level.

\subsection{{\Ae} subclass}

\begin{figure*}[h]
\resizebox{6cm}{!}{\includegraphics{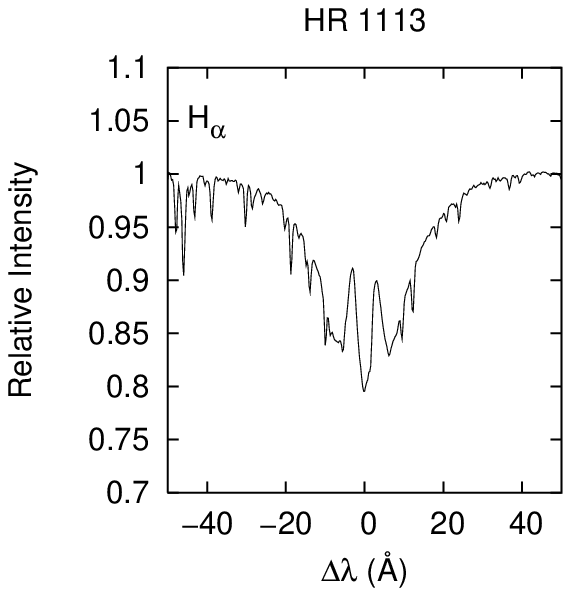}}
\resizebox{6cm}{!}{\includegraphics{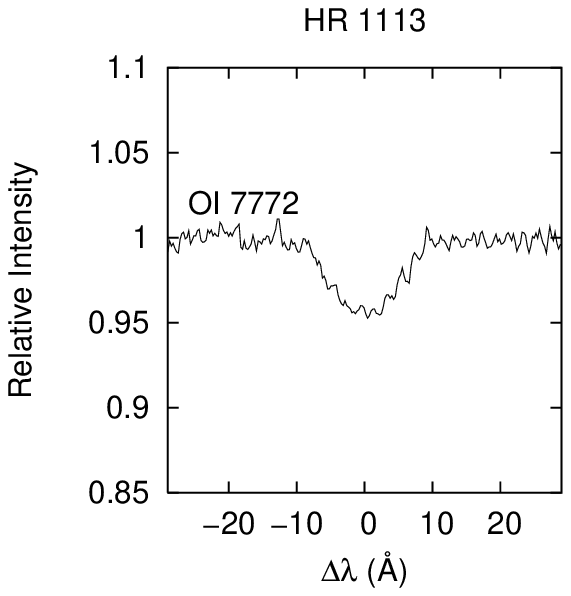}}
\caption{Profiles of {\Halpha} and
{\oxir}
lines of \object{HR~1113}.}
\label{HR1113}
\end{figure*}

\begin{figure*}[h]
\resizebox{6cm}{!}{\includegraphics{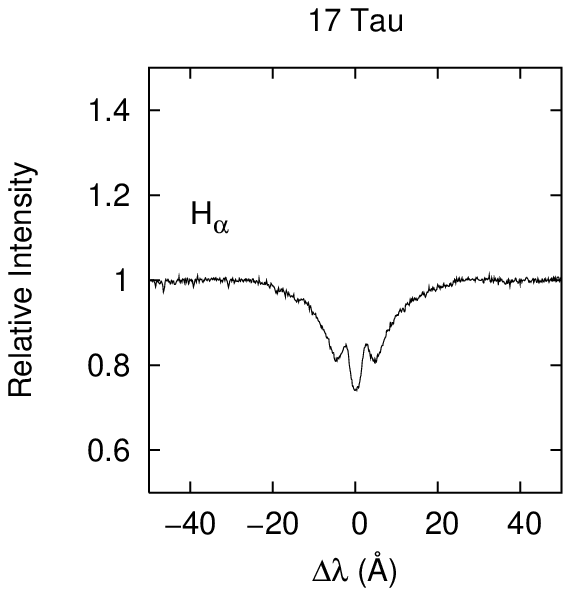}}
\resizebox{6cm}{!}{\includegraphics{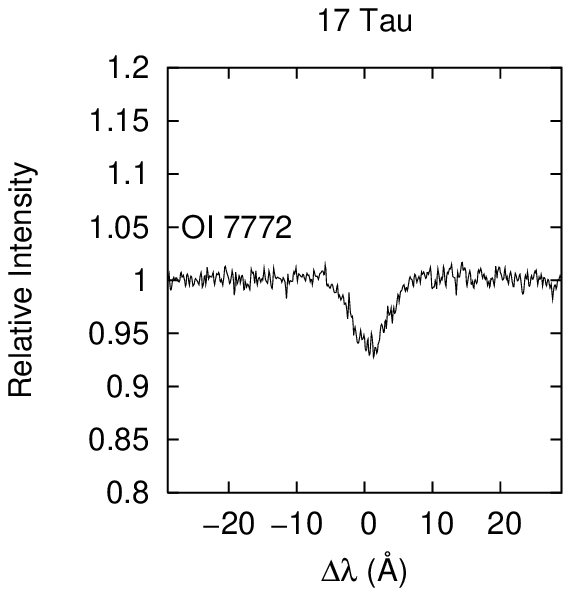}}
\caption{Profiles of {\Halpha} and {\oxir} lines of \object{17~Tau}.}
\label{17tau}
\end{figure*}

\begin{figure*}[h]
\resizebox{6cm}{!}{\includegraphics{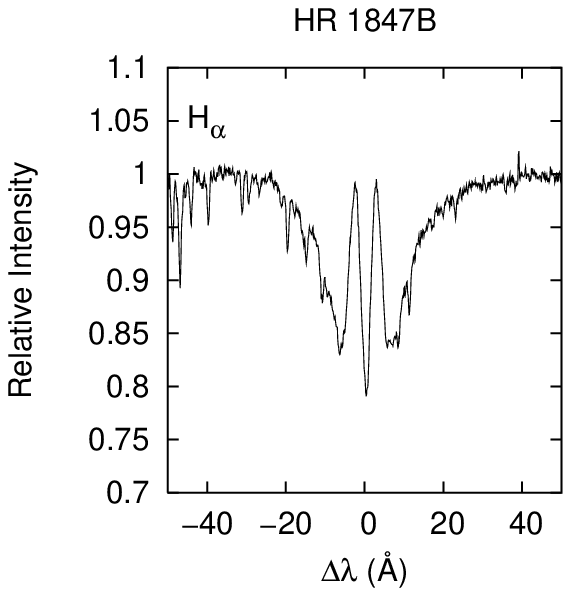}}
\resizebox{6cm}{!}{\includegraphics{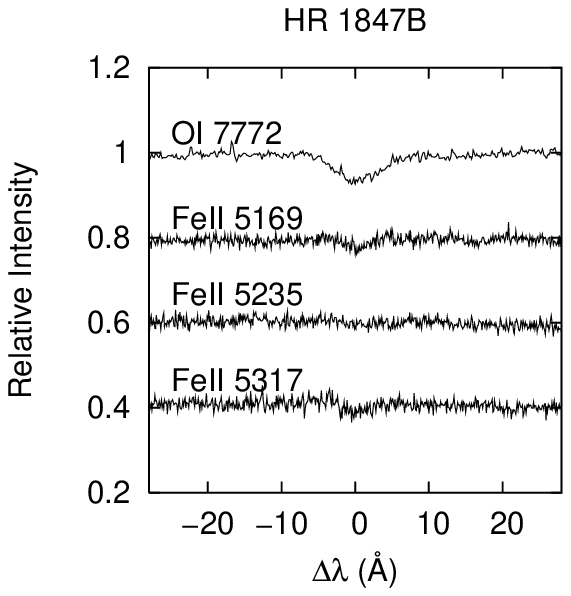}}
\caption{Profiles of {\Halpha}, {\feii} 5169, 5235, 5317\,{\AA}, and
{\oxir} lines of \object{HR~1847B}.}
\label{HR1847B}
\end{figure*}

\begin{figure*}[h]
\resizebox{6cm}{!}{\includegraphics{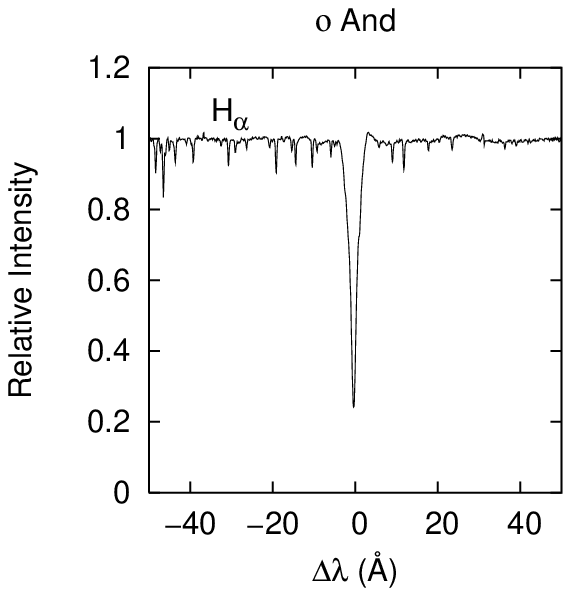}}
\resizebox{6cm}{!}{\includegraphics{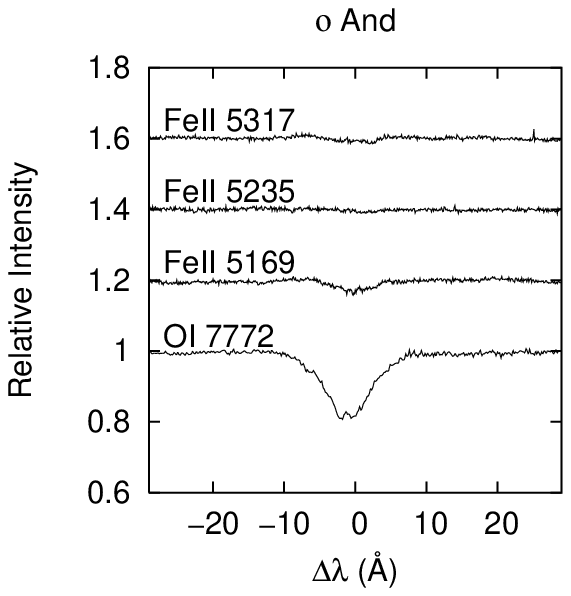}}
\caption{Profiles of {\Halpha}, {\feii} 5169, 5235, 5317\,{\AA},
and {\oxir} lines of \object{$o$~And}.}
\label{omiand}
\end{figure*}

\clearpage

\subsection{{\Ea} subclass}

\begin{figure*}[ht]
\resizebox{6cm}{!}{\includegraphics{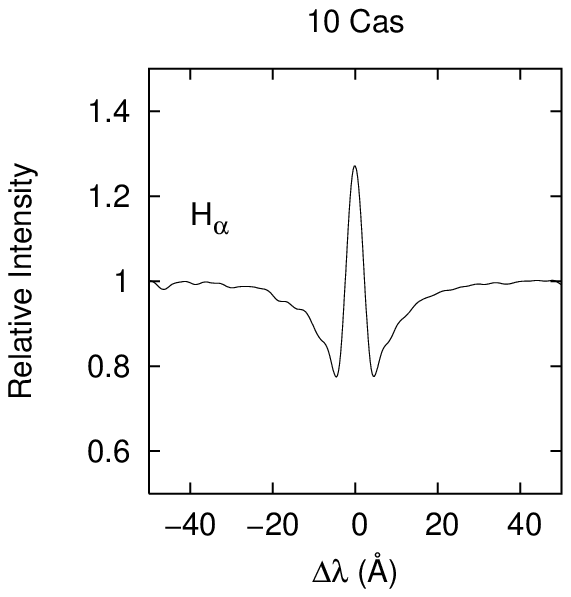}}
\resizebox{6cm}{!}{\includegraphics{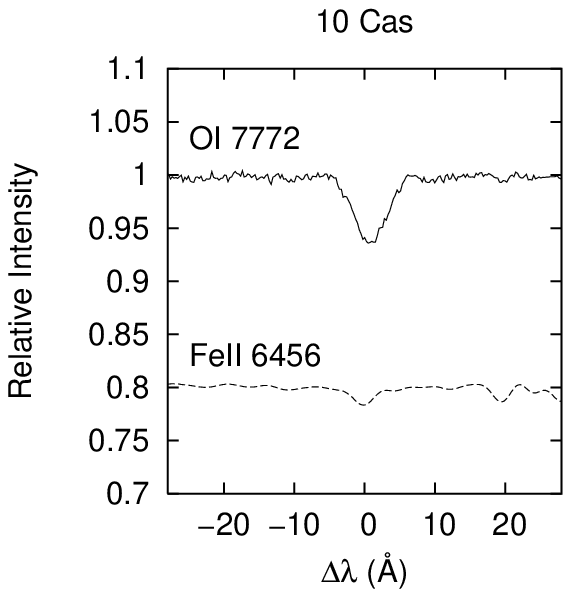}}
\caption{Profiles of {\Halpha}, {\feii} 6456\,{\AA}, and {\oxir} lines
of \object{10~Cas}.}
\label{10cas}
\end{figure*}

\begin{figure*}[ht]
\resizebox{6cm}{!}{\includegraphics{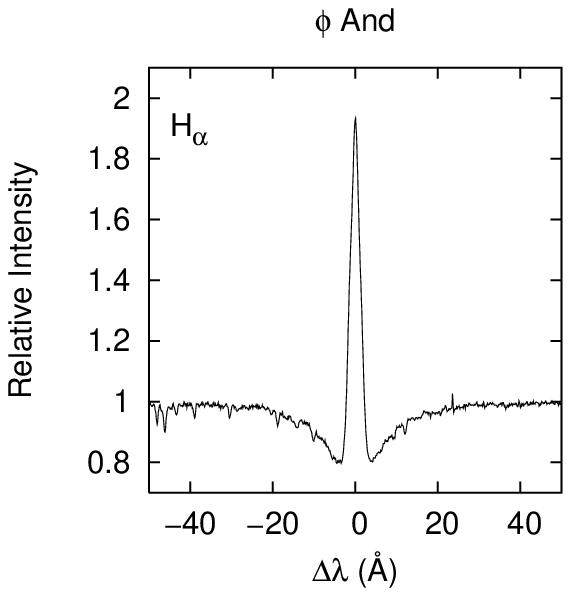}}
\resizebox{6cm}{!}{\includegraphics{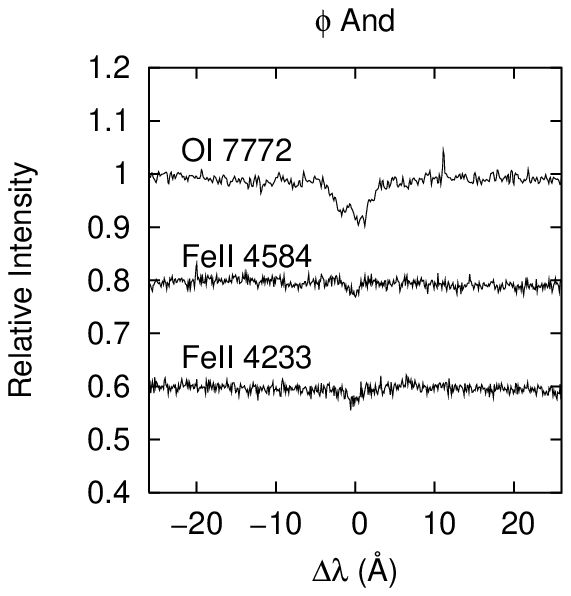}}
\caption{Profiles of {\Halpha},
{\feii} 4233, 4584\,{\AA},
 and {\oxir} lines of
\object{$\phi$~And}.}
\label{phiand}
\end{figure*}

\begin{figure*}[ht]
\resizebox{6cm}{!}{\includegraphics{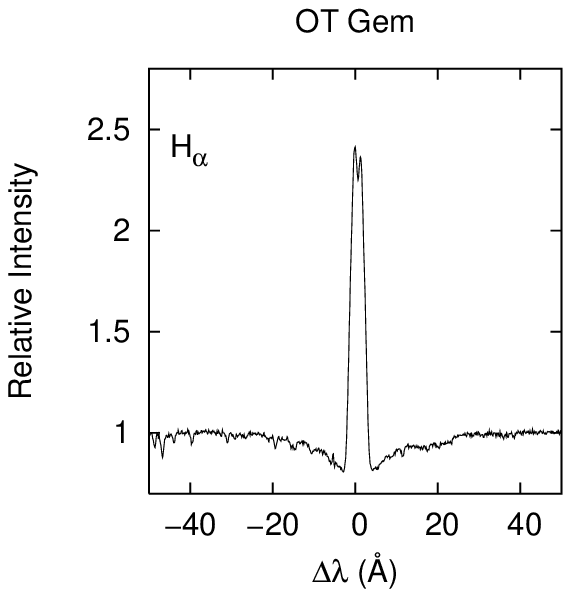}}
\resizebox{6cm}{!}{\includegraphics{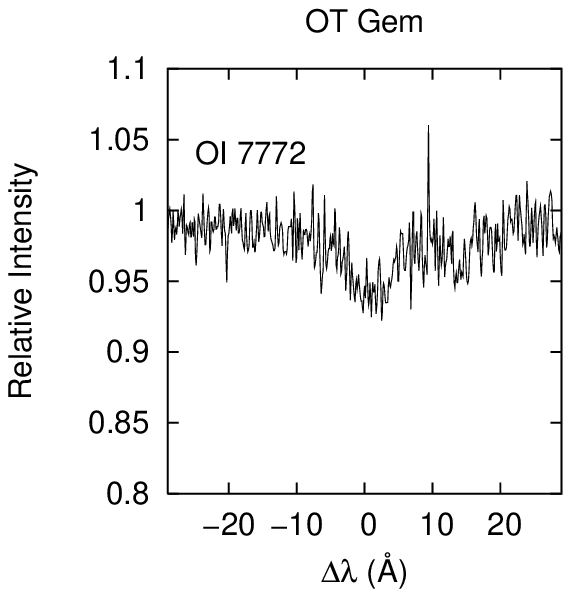}}
\caption{Profiles of {\Halpha} and {\oxir} lines of \object{OT~Gem}.}
\label{otgem}
\end{figure*}

\begin{figure*}[ht]
\resizebox{6cm}{!}{\includegraphics{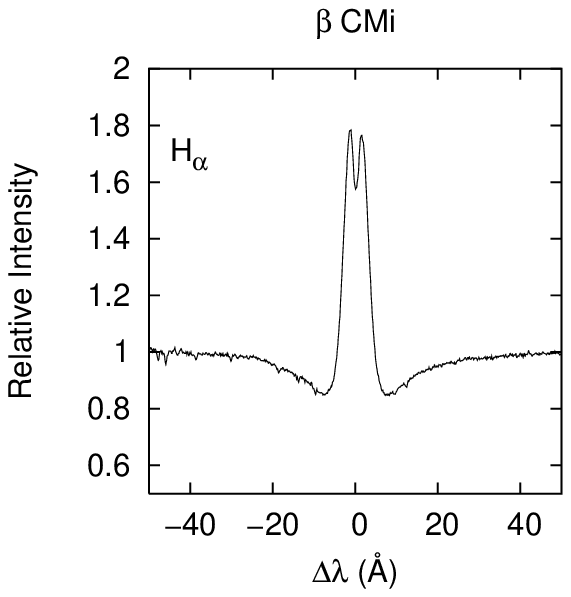}}
\resizebox{6cm}{!}{\includegraphics{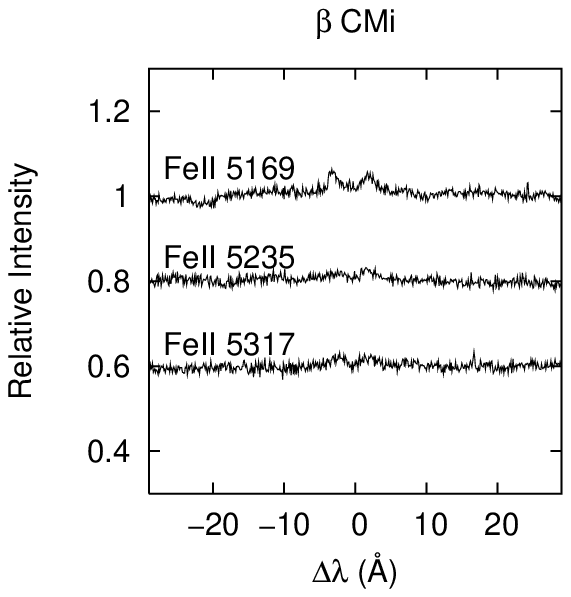}}
\resizebox{6cm}{!}{\includegraphics{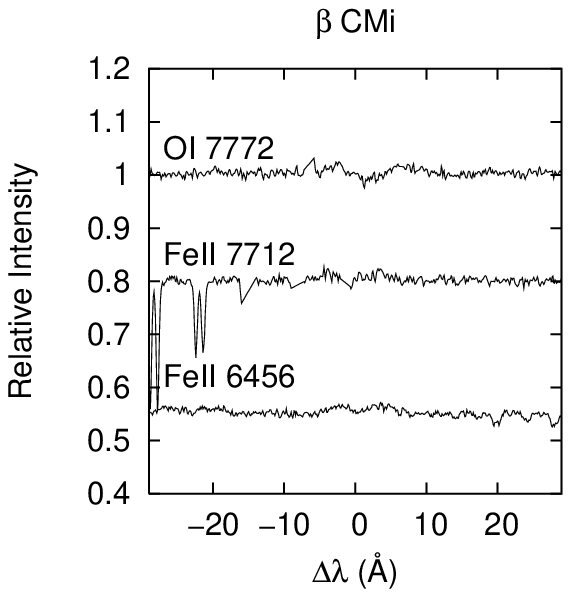}}
\caption{Profiles of {\Halpha}, {\feii} 5169, 5235, 5317,
6456,
7712,
and {\oxir} lines of \object{$\beta$~CMi}.}
\label{betcmi}
\end{figure*}

\begin{figure*}[ht]
\resizebox{6cm}{!}{\includegraphics{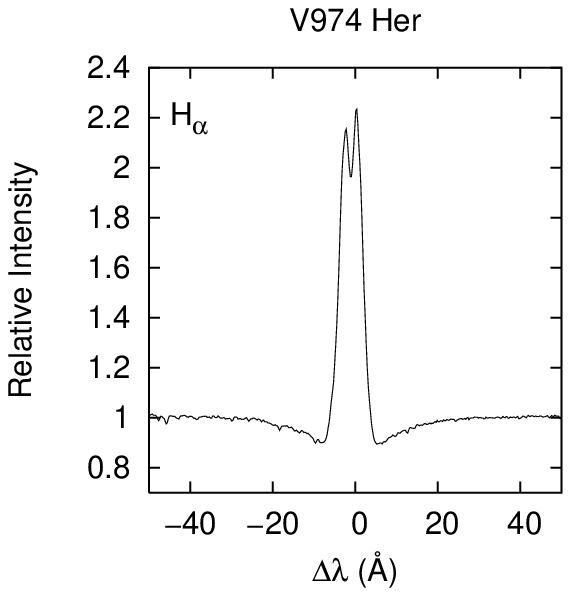}}
\resizebox{6cm}{!}{\includegraphics{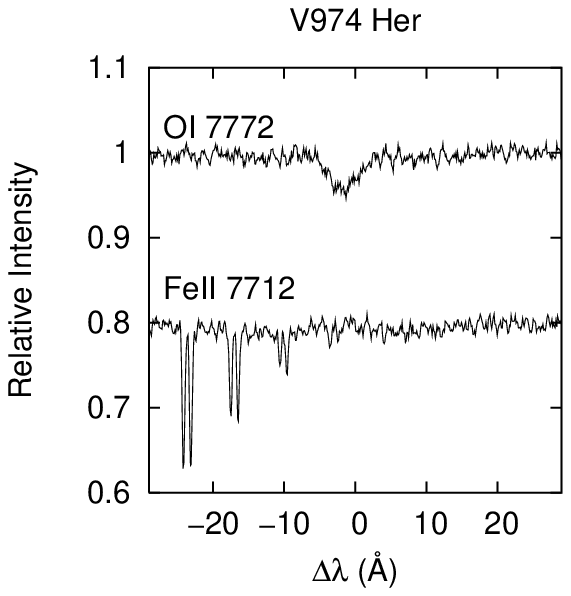}}
\caption{Profiles of {\Halpha}, {\feii} 77712\,{\AA}, and {\oxir} lines
of \object{V974\,Her} (\object{HD~164447}).}
\label{v974her}
\end{figure*}

\begin{figure*}[ht]
\resizebox{6cm}{!}{\includegraphics{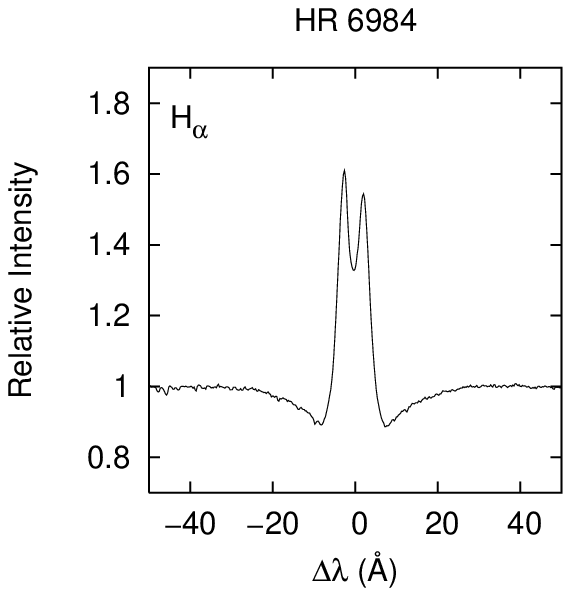}}
\resizebox{6cm}{!}{\includegraphics{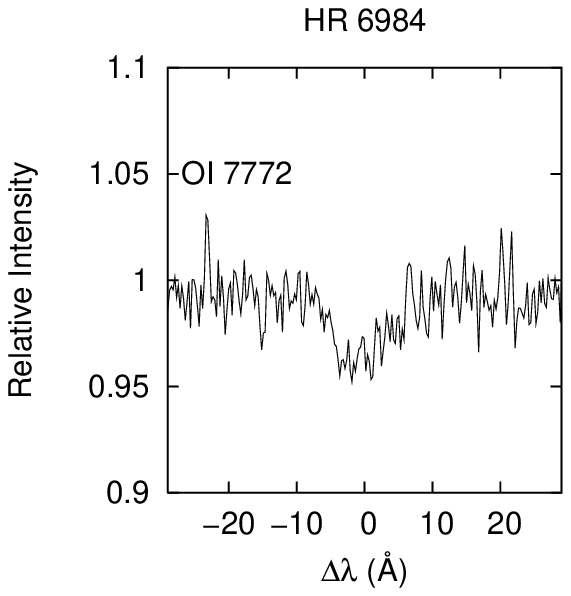}}
\caption{Profiles of {\Halpha} and {\oxir} lines of
\object{HR~6984} (\object{HD~171780}).}
\label{HR6984}
\end{figure*}

\begin{figure*}[ht]
\resizebox{6cm}{!}{\includegraphics{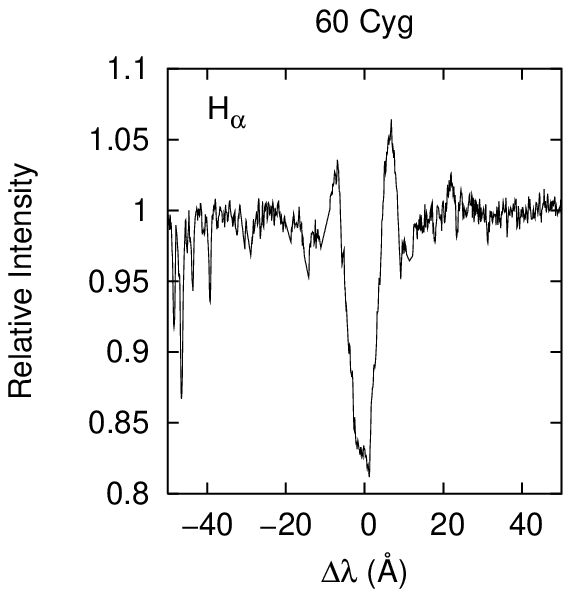}}
\resizebox{6cm}{!}{\includegraphics{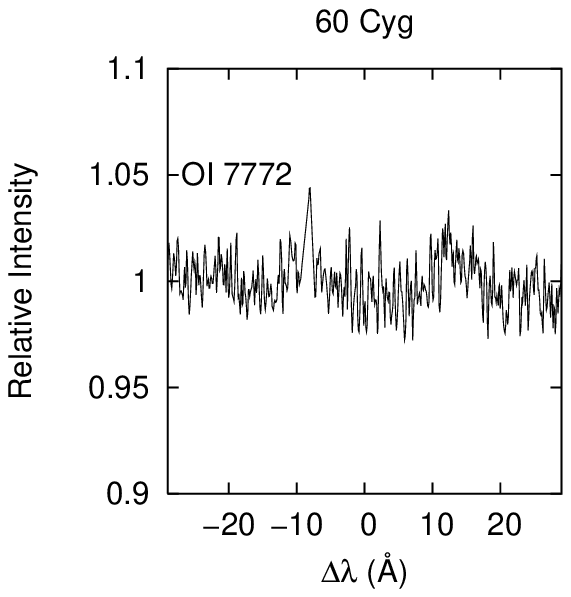}}
\caption{Profiles of {\Halpha} and {\oxir} lines of \object{60~Cyg}.}
\label{60cyg}
\end{figure*}

\begin{figure*}[ht]
\resizebox{6cm}{!}{\includegraphics{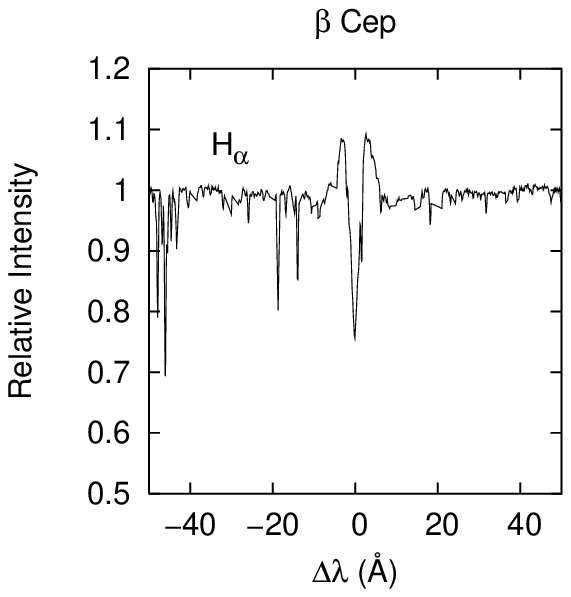}}
\resizebox{6cm}{!}{\includegraphics{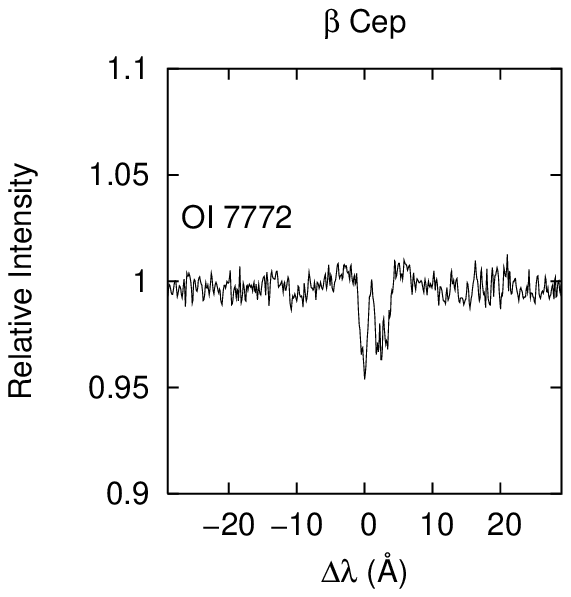}}
\caption{Profiles of {\Halpha} and {\oxir} lines of
\object{$\beta$~Cep}.}
\label{betcep}
\end{figure*}

\begin{figure*}[ht]
\resizebox{6cm}{!}{\includegraphics{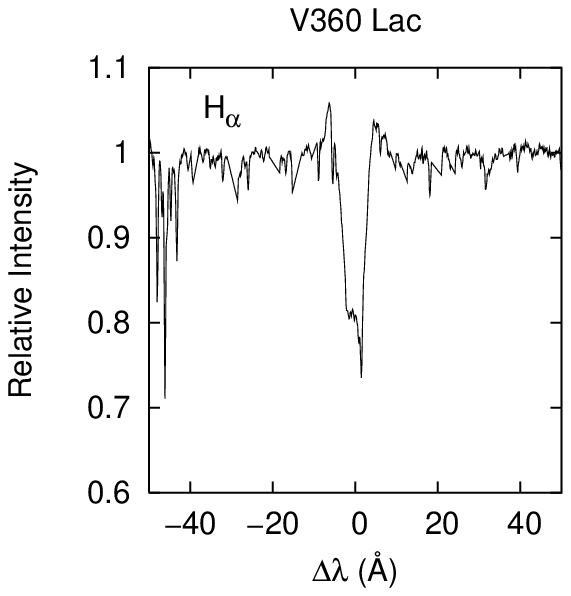}}
\resizebox{6cm}{!}{\includegraphics{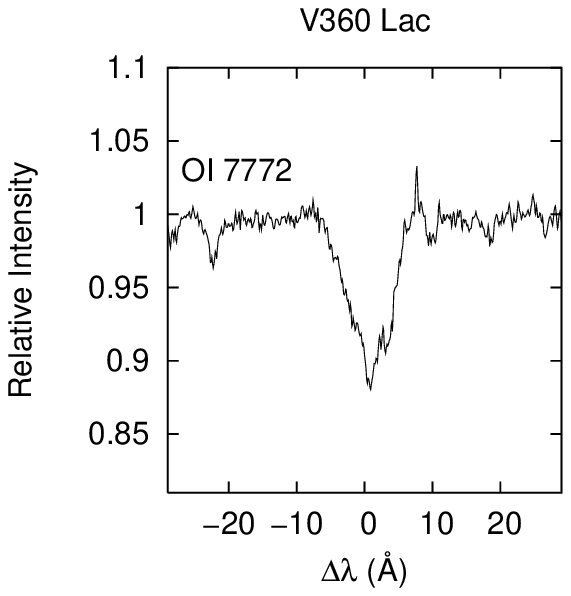}}
\caption{Profiles of {\Halpha} and {\oxir} lines of
\object{V~360~Lac}.}
\label{V360lac}
\end{figure*}

\clearpage

\subsection{{\Em} subclass}

\begin{figure*}[th]
\resizebox{6cm}{!}{\includegraphics{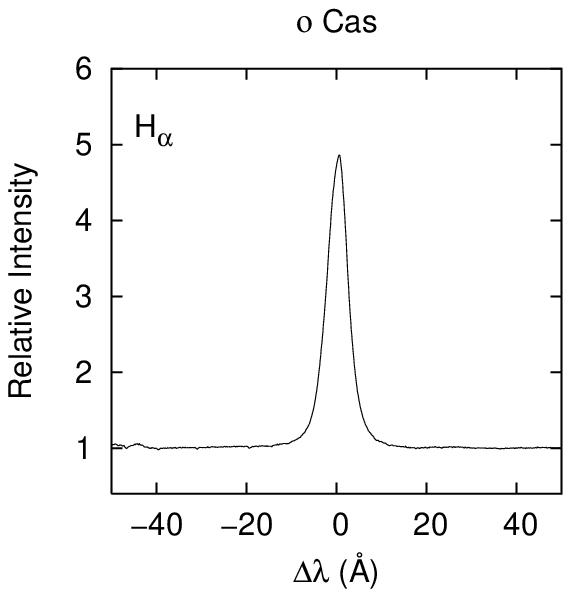}}
\resizebox{6cm}{!}{\includegraphics{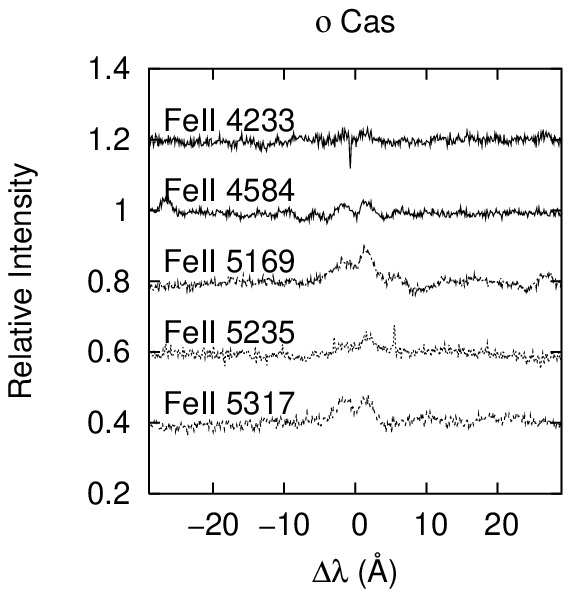}}
\resizebox{6cm}{!}{\includegraphics{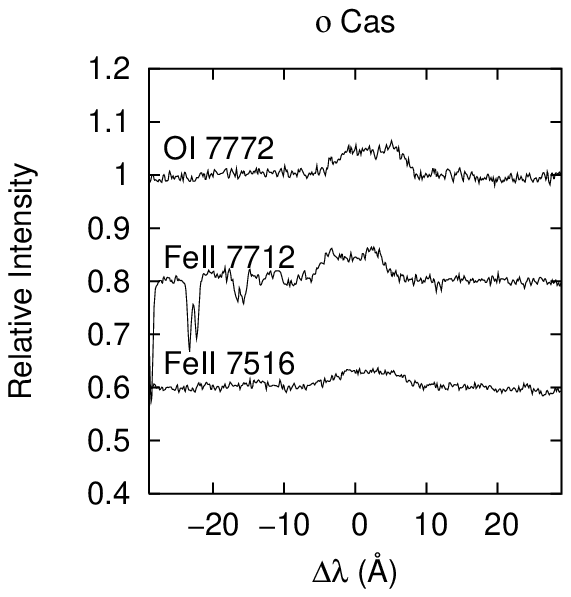}}
\caption{Profiles of {\Halpha}, {\feii}
4233, 4584,
5169, 5235, 5317, 7516, 7712\,{\AA},
and {\oxir} lines of \object{$o$~Cas}.}
\label{omicas}
\end{figure*}

\begin{figure*}[ht]
\resizebox{6cm}{!}{\includegraphics{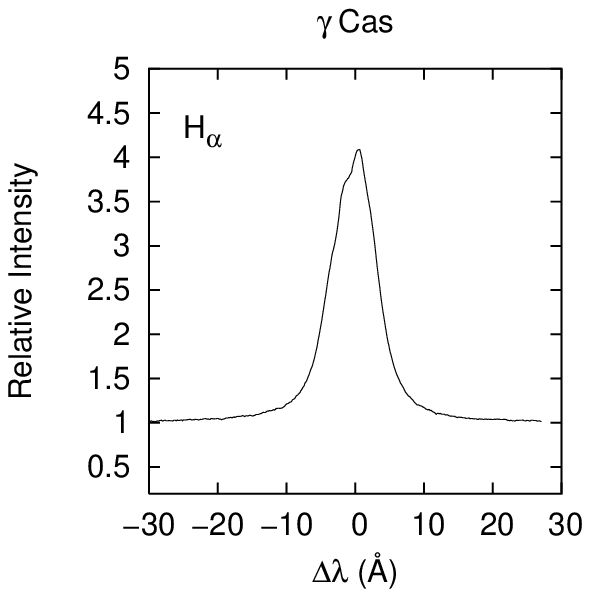}}
\resizebox{6cm}{!}{\includegraphics{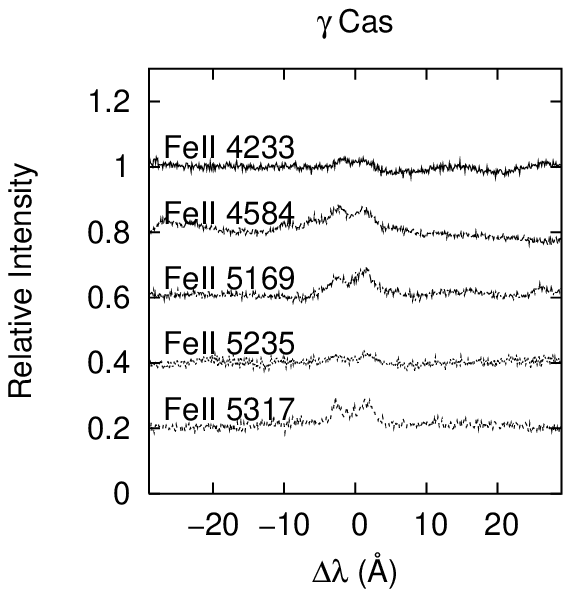}}
\resizebox{6cm}{!}{\includegraphics{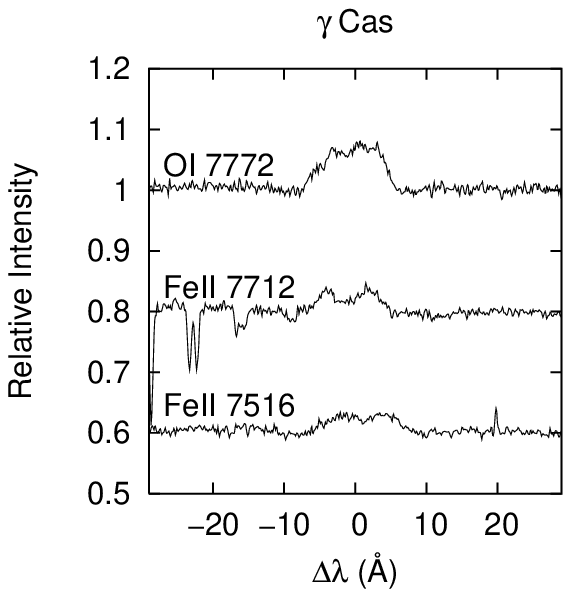}}
\caption{Profiles of {\Halpha}, {\feii}
4233, 4584,
5169, 5235, 5317, 7516, 7712\,{\AA},
and {\oxir} lines of \object{$\gamma$~Cas}.}
\label{gamcas}
\end{figure*}

\begin{figure*}[ht]
\resizebox{6cm}{!}{\includegraphics{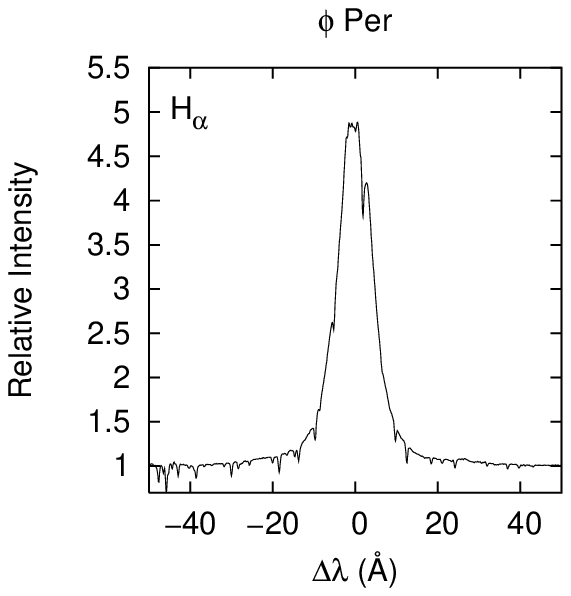}}
\resizebox{6cm}{!}{\includegraphics{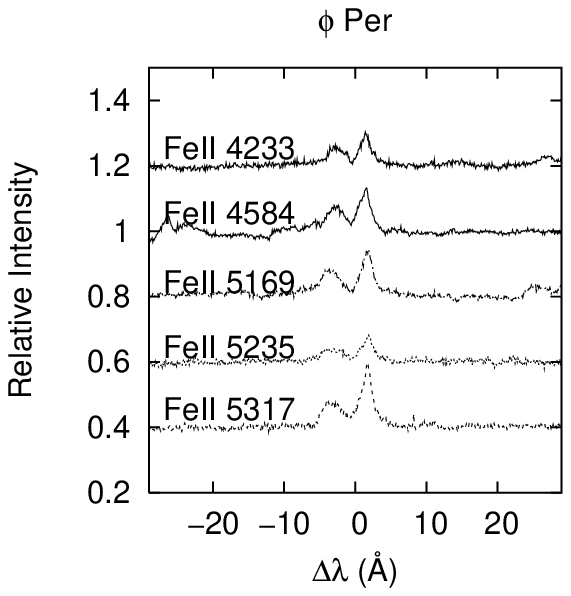}}
\resizebox{6cm}{!}{\includegraphics{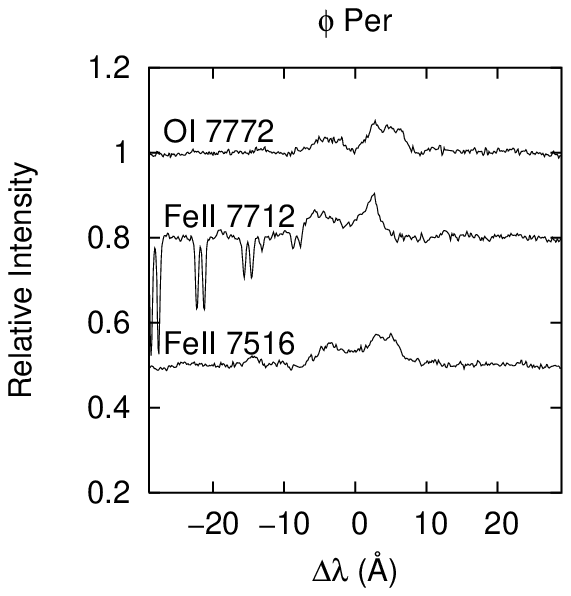}}
\caption{Profiles of {\Halpha}, {\feii}
4233, 4584,
5169, 5235, 5317, 7516, 7712\,{\AA},
and {\oxir} lines of \object{$\phi$~Per}.}
\label{phiper}
\end{figure*}

\begin{figure*}[ht]
\resizebox{6cm}{!}{\includegraphics{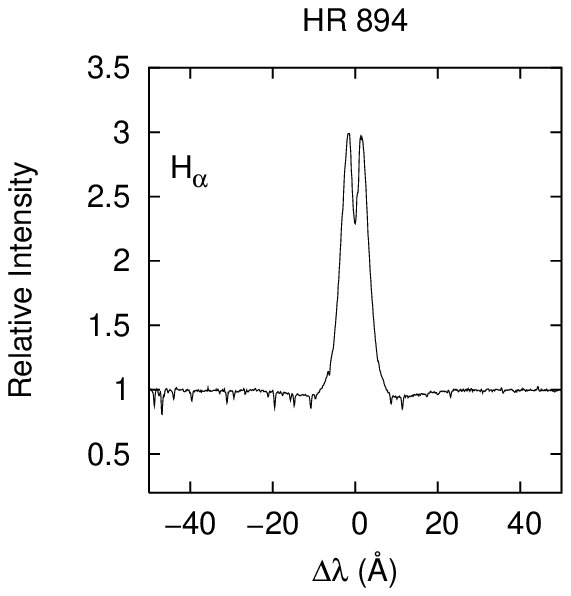}}
\resizebox{6cm}{!}{\includegraphics{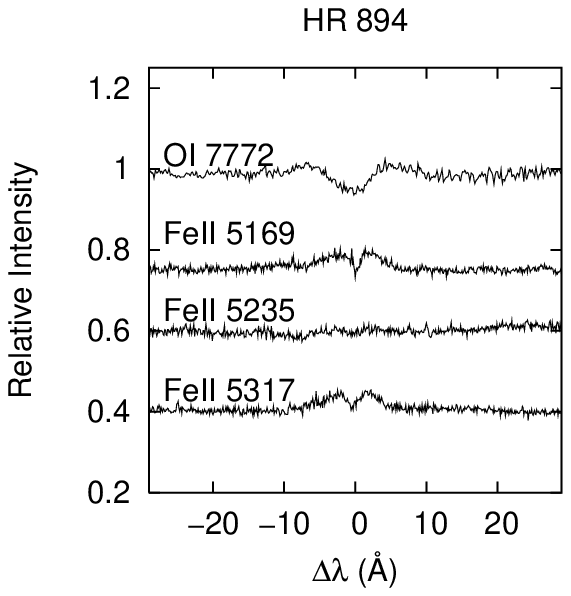}}
\caption{Profiles of {\Halpha}, {\feii} 5169, 5235, 5317\,{\AA}, and
{\oxir} lines of \object{HR~894}.}
\label{HR894}
\end{figure*}

\begin{figure*}[ht]
\resizebox{6cm}{!}{\includegraphics{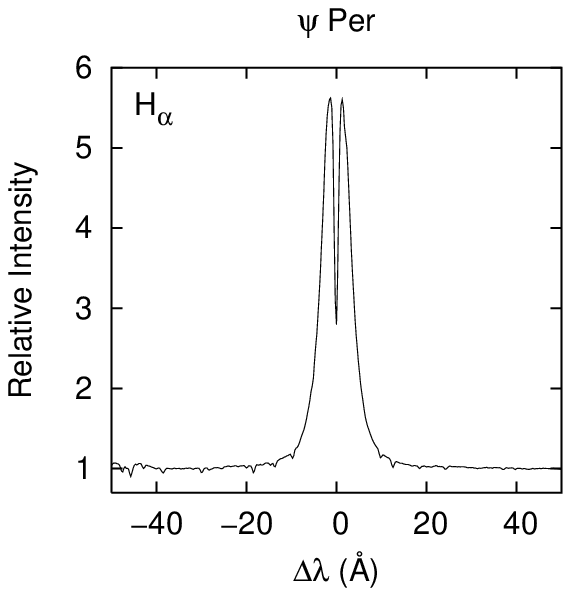}}
\resizebox{6cm}{!}{\includegraphics{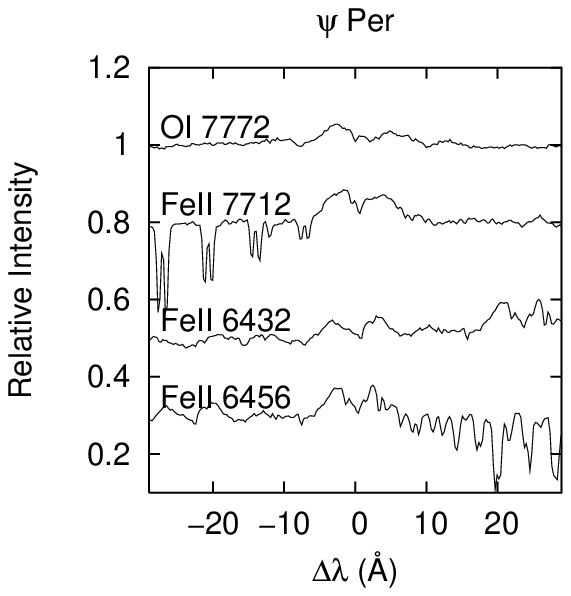}}
\caption{Profiles of {\Halpha} and {\feii} 6432, 6456,
7712\,{\AA}
lines
of \object{$\psi$~Per}.}
\label{psiper}
\end{figure*}

\begin{figure*}[ht]
\resizebox{6cm}{!}{\includegraphics{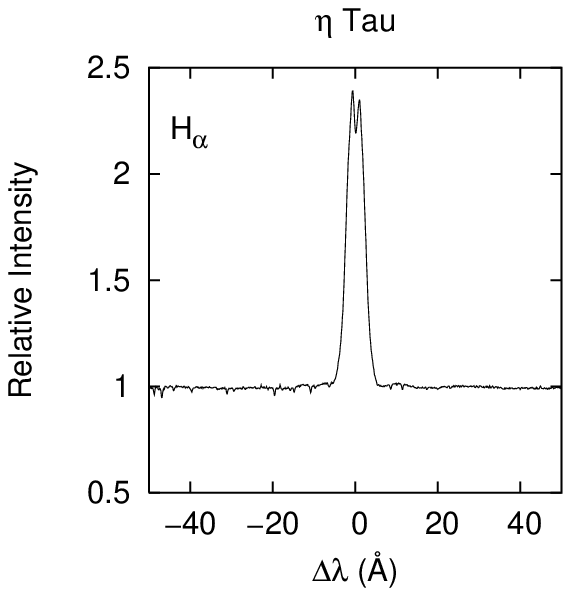}}
\resizebox{6cm}{!}{\includegraphics{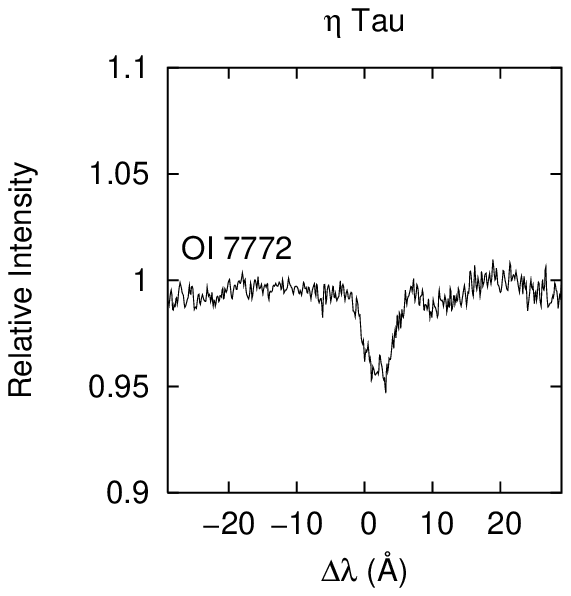}}
\caption{Profiles of {\Halpha} and {\oxir} lines of
\object{$\eta$~Tau}.}
\label{etatau}
\end{figure*}

\begin{figure*}[ht]
\resizebox{6cm}{!}{\includegraphics{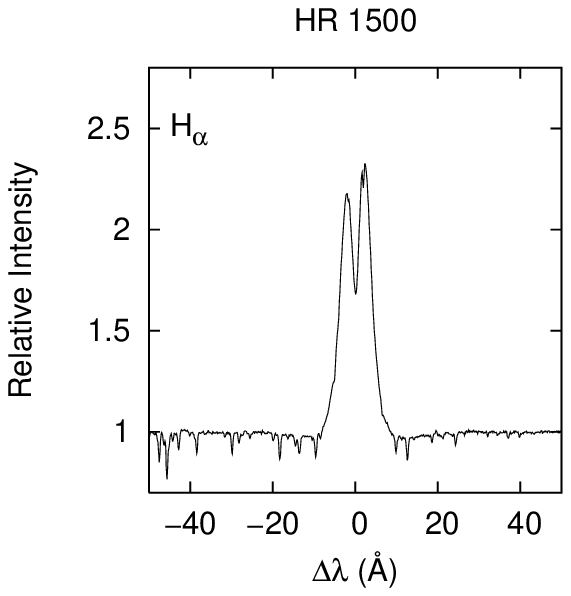}}
\resizebox{6cm}{!}{\includegraphics{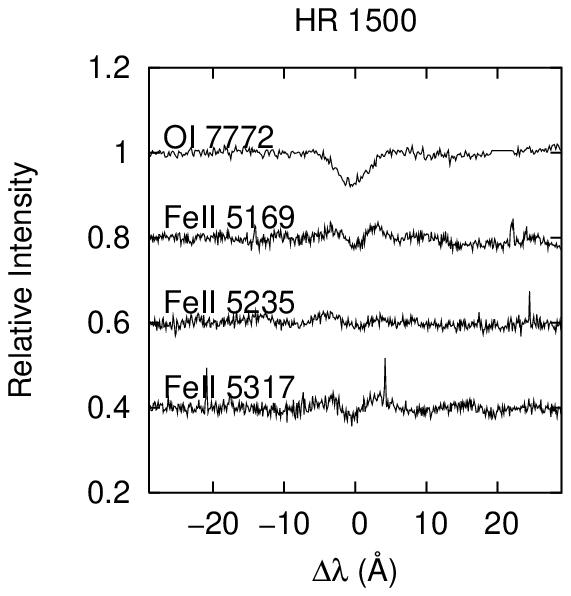}}
\caption{Profiles of {\Halpha}, {\feii} 5169, 5235, 5317\,{\AA}, and
{\oxir} lines of \object{HR~1500}.}
\label{HR1500}
\end{figure*}

\begin{figure*}[ht]
\resizebox{6cm}{!}{\includegraphics{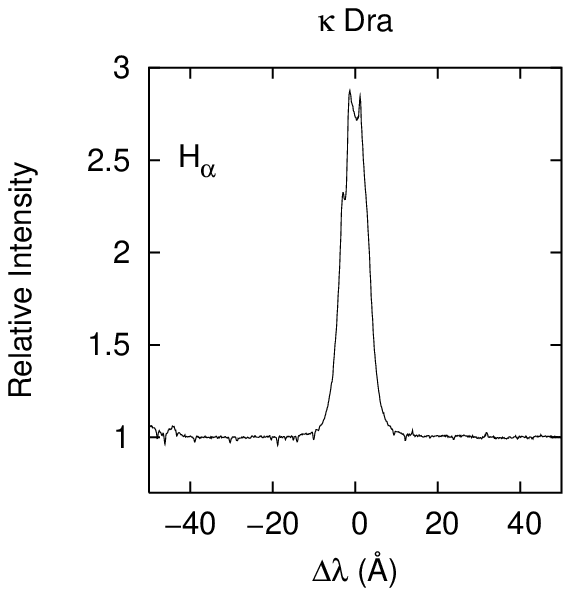}}
\resizebox{6cm}{!}{\includegraphics{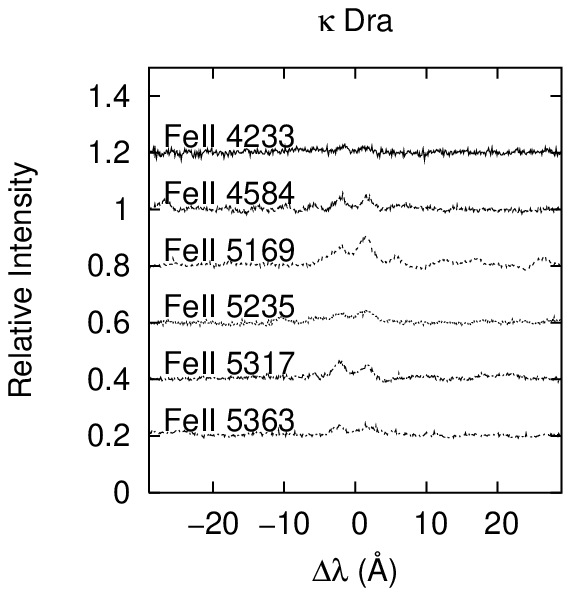}}
\resizebox{6cm}{!}{\includegraphics{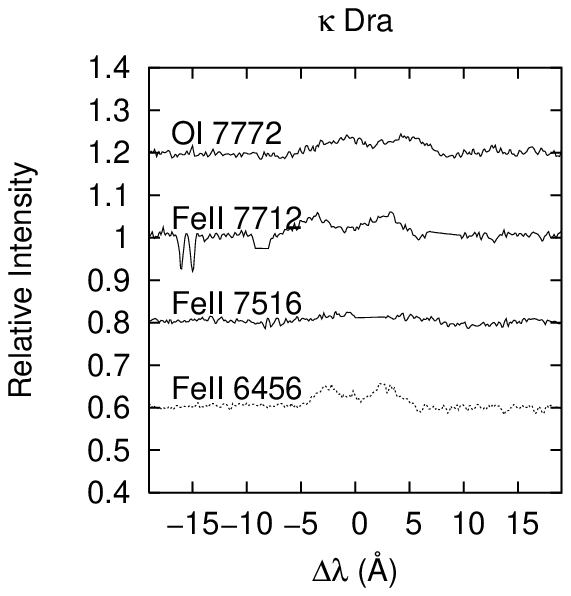}}
\caption{Profiles of {\Halpha}, {\feii}
4233, 4584,
5169, 5235, 5317, 5363, 6456,
7516, 7712\,{\AA}, and {\oxir} lines of \object{$\kappa$~Dra}.}
\label{kdra}
\end{figure*}

\begin{figure*}[ht]
\resizebox{6cm}{!}{\includegraphics{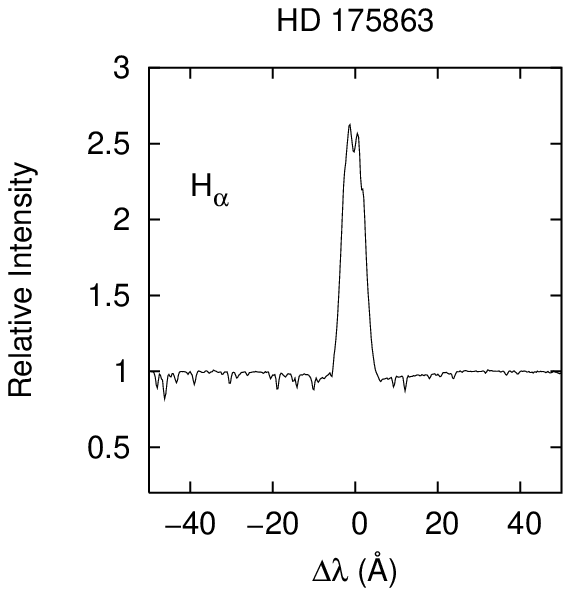}}
\resizebox{6cm}{!}{\includegraphics{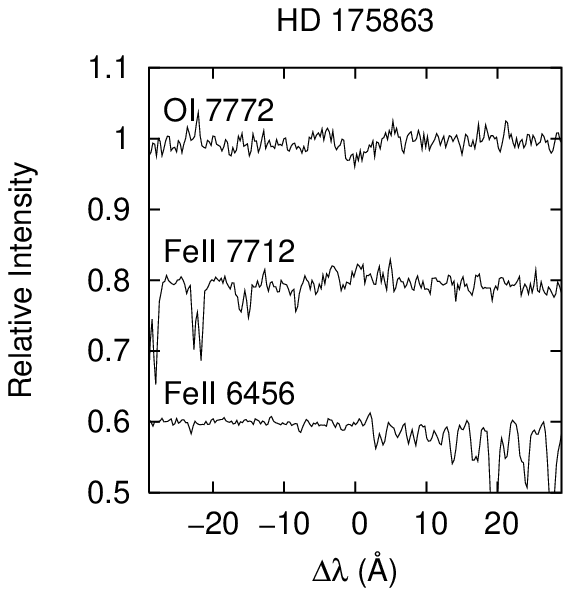}}
\caption{Profiles of {\Halpha}, {\feii}
6456, 7712\,{\AA},
 and {\oxir} lines of
\object{HD~175863}.}
\label{HD175863}
\end{figure*}

\begin{figure*}[ht]
\resizebox{6cm}{!}{\includegraphics{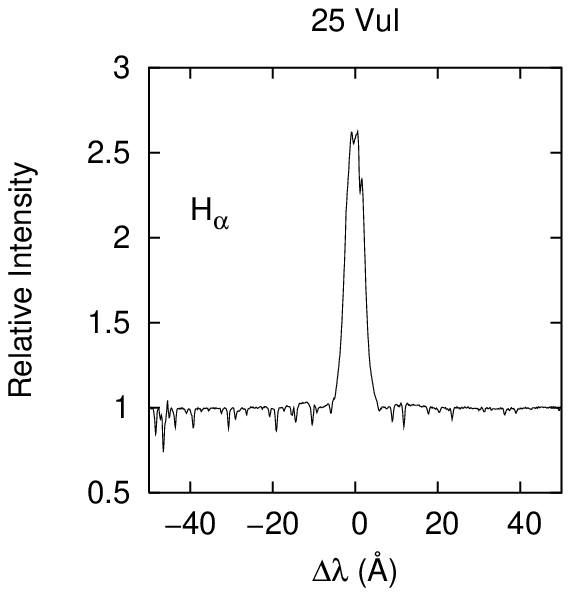}}
\resizebox{6cm}{!}{\includegraphics{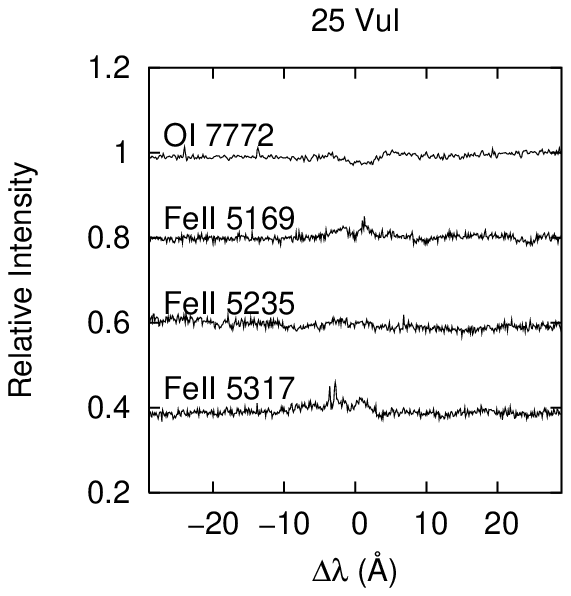}}
\caption{Profiles of {\Halpha}, {\feii} 5169, 5235, 5317\,{\AA}, and
{\oxir} lines of \object{25~Vul}.}
\label{25vul}
\end{figure*}

\begin{figure*}[ht]
\resizebox{6cm}{!}{\includegraphics{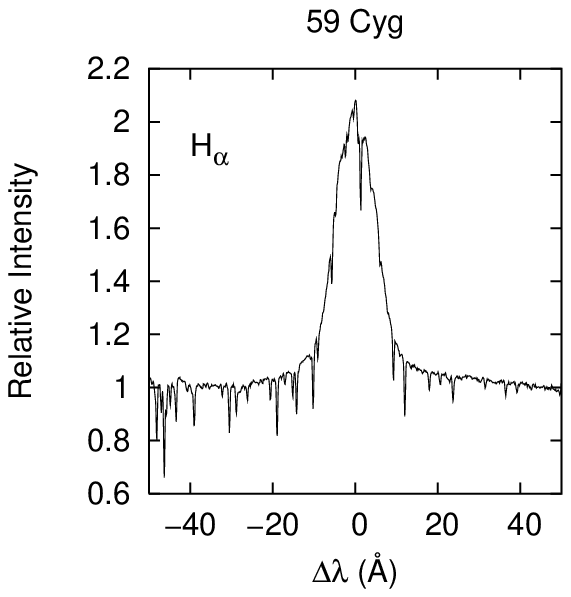}}
\resizebox{6cm}{!}{\includegraphics{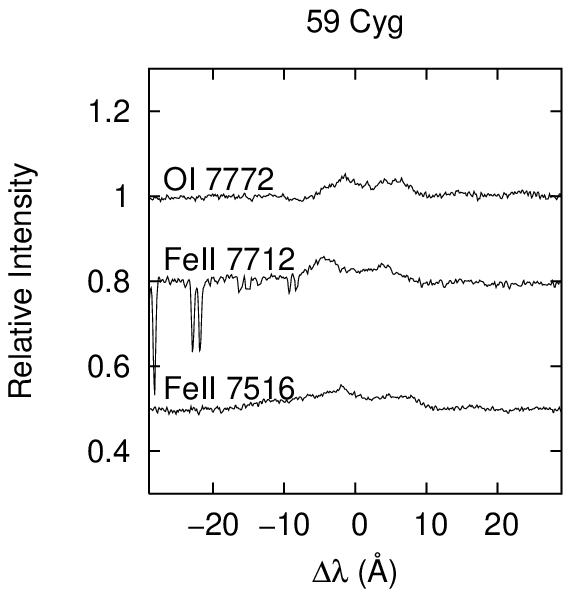}}
\caption{Profiles of {\Halpha}, {\feii} 7516, 7712\,{\AA}, and {\oxir}
lines of \object{59~Cyg}.}
\label{59cyg}
\end{figure*}

\begin{figure*}[ht]
\resizebox{6cm}{!}{\includegraphics{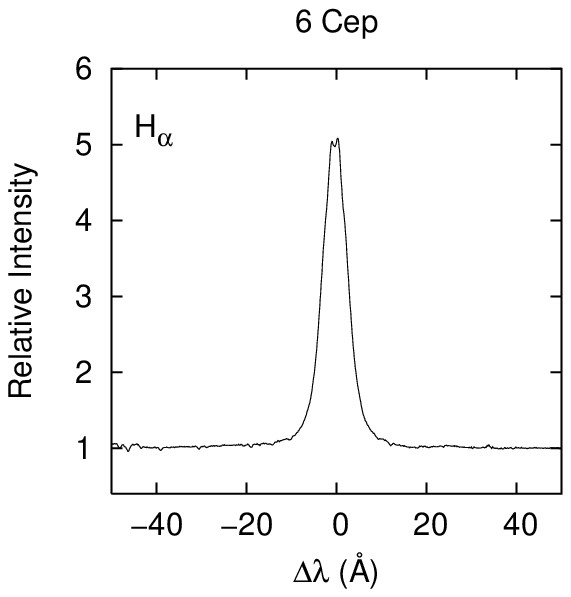}}
\resizebox{6cm}{!}{\includegraphics{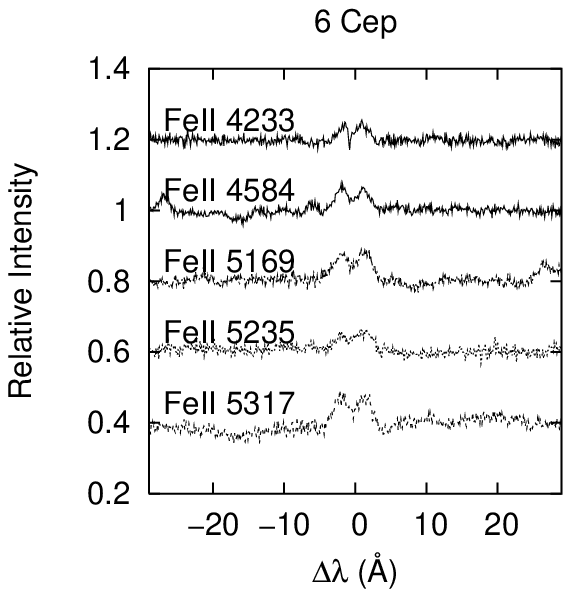}}
\resizebox{6cm}{!}{\includegraphics{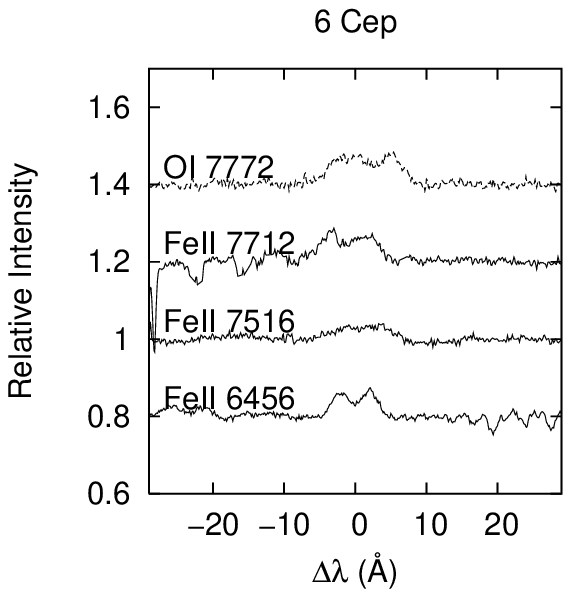}}
\caption{Profiles of {\Halpha}, {\feii}
4233, 4584,
5169, 5235, 5317, 6456,
7516, 7712\,{\AA}, and {\oxir} lines of \object{6~Cep}.}
\label{6cep}
\end{figure*}

\begin{figure*}[ht]
\resizebox{6cm}{!}{\includegraphics{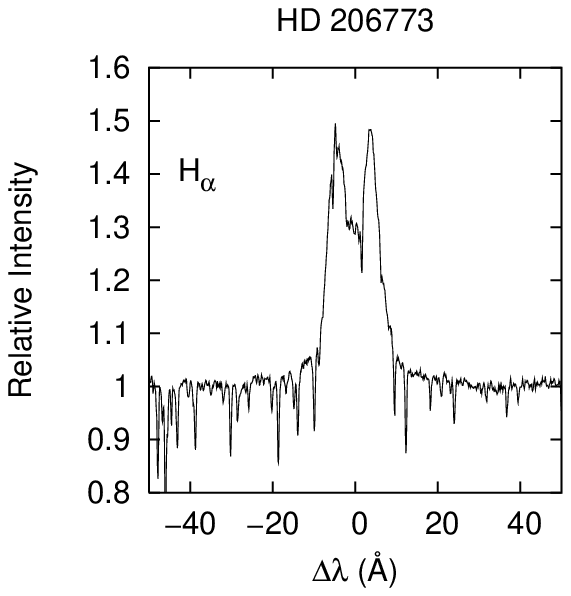}}
\resizebox{6cm}{!}{\includegraphics{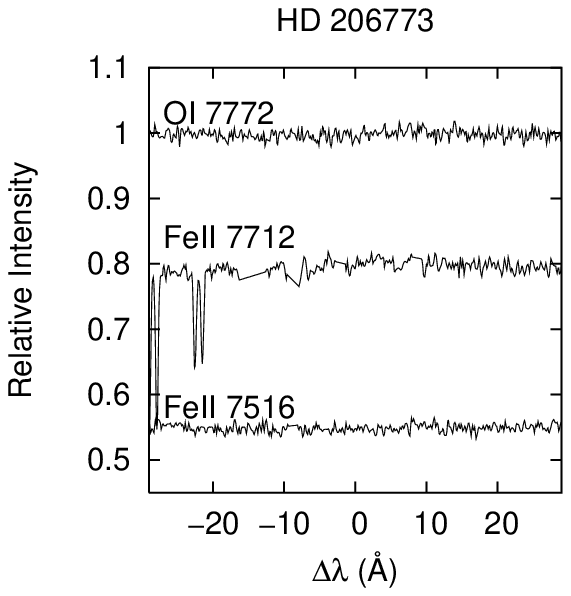}}
\caption{Profiles of {\Halpha}, {\feii} 7516,
7712\,{\AA}, and {\oxir}
lines of \object{HD~206773}.}
\label{HD206773}
\end{figure*}

\begin{figure*}[ht]
\resizebox{6cm}{!}{\includegraphics{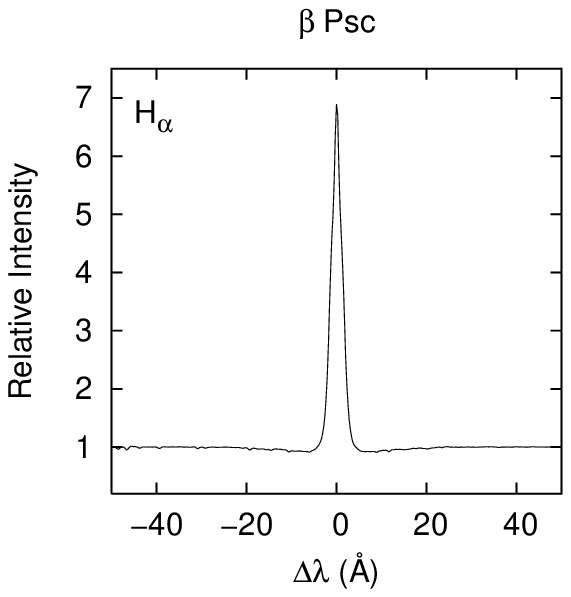}}
\resizebox{6cm}{!}{\includegraphics{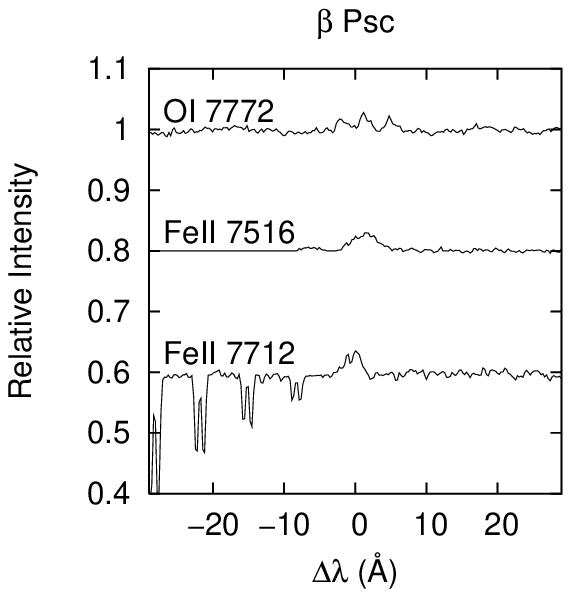}}
\caption{Profiles of {\Halpha},
{\feii} 7516, 7712\,{\AA},
and {\oxir} lines of
\object{$\beta$~Psc}.}
\label{betpsc}
\end{figure*}

\begin{figure*}[ht]
\resizebox{6cm}{!}{\includegraphics{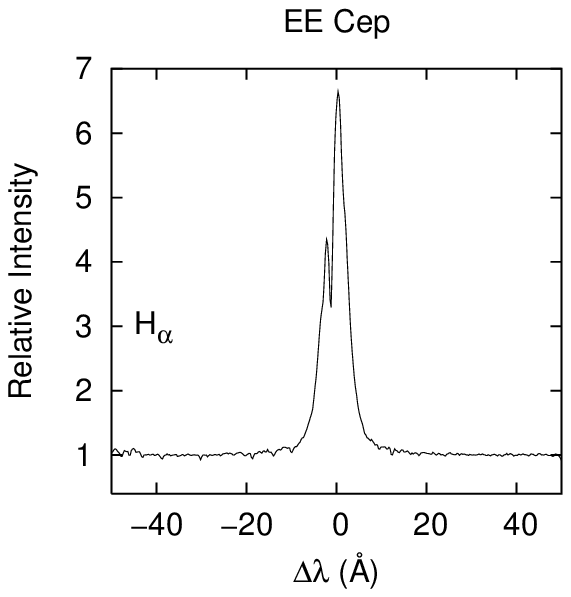}}
\resizebox{6cm}{!}{\includegraphics{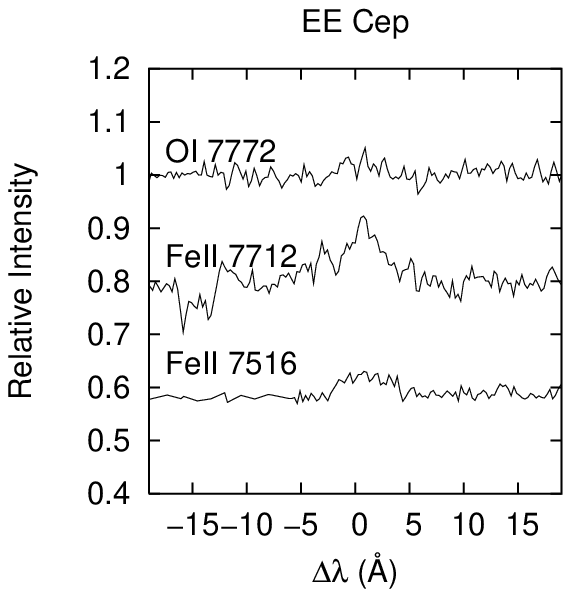}}
\caption{Profiles of {\Halpha},
{\feii} 7516, 7712\,{\AA},
and {\oxir} lines of
\object{EE\,Cep}.}
\label{eecep}
\end{figure*}

\clearpage

\subsection{{\Sh} subclass}

\begin{figure*}[ht]
\resizebox{6cm}{!}{\includegraphics{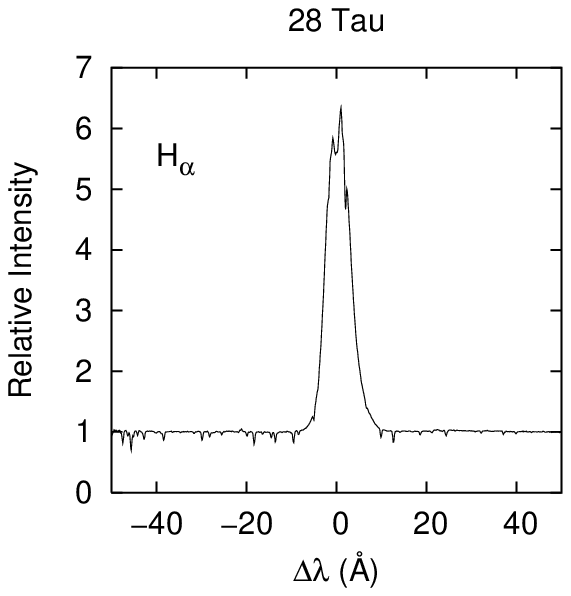}}
\resizebox{6cm}{!}{\includegraphics{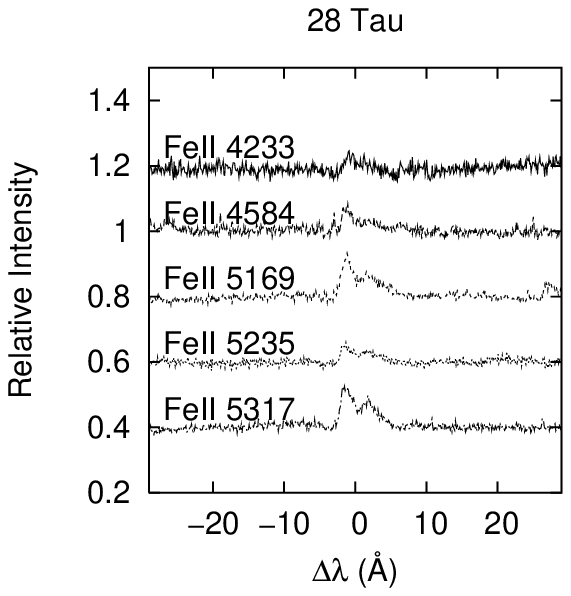}}
\resizebox{6cm}{!}{\includegraphics{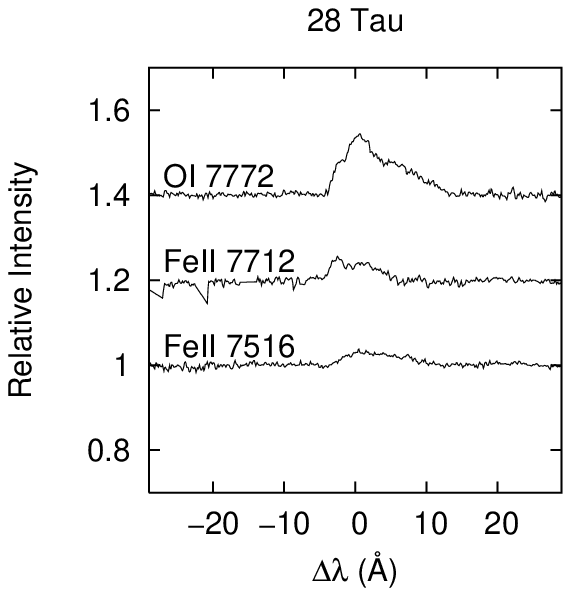}}
\caption{Profiles of {\Halpha}, {\feii}
4233, 4584,
5169, 5235, 5317, 7516, 7712\,{\AA},
and {\oxir} lines of \object{28~Tau}.}
\label{butau}
\end{figure*}

\begin{figure*}[ht]
\resizebox{6cm}{!}{\includegraphics{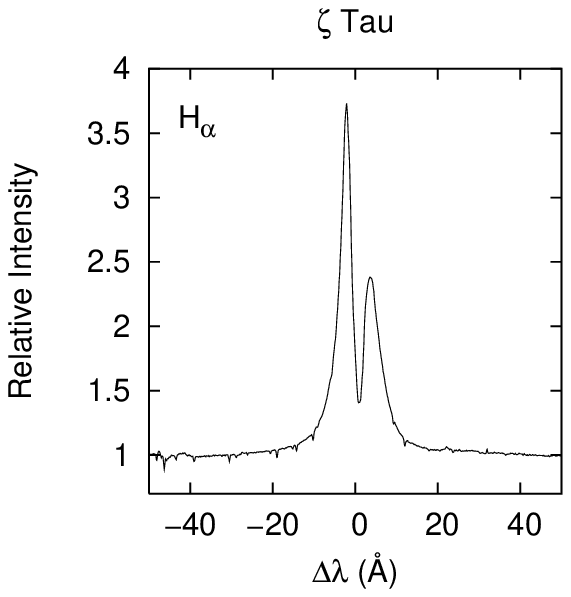}}
\resizebox{6cm}{!}{\includegraphics{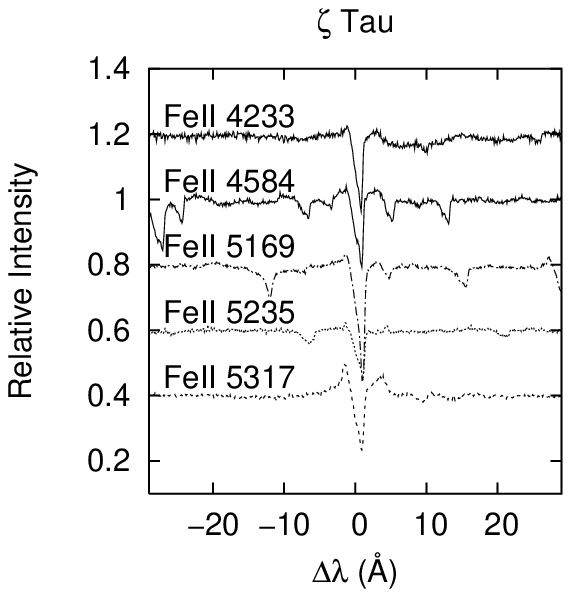}}
\resizebox{6cm}{!}{\includegraphics{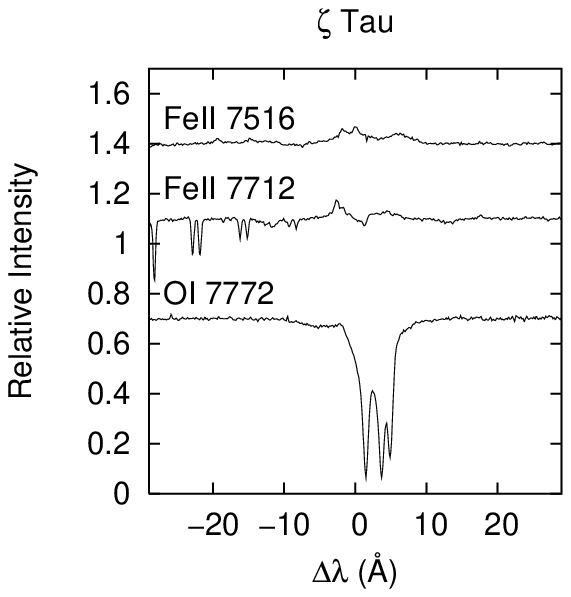}}
\caption{Profiles of {\Halpha}, {\feii} 4233, 4584,
5169, 5235, 5317,
7516, 7712\,{\AA},
and {\oxir} lines of \object{$\zeta$~Tau}.}
\label{zettau}
\end{figure*}

\begin{figure*}[ht]
\resizebox{6cm}{!}{\includegraphics{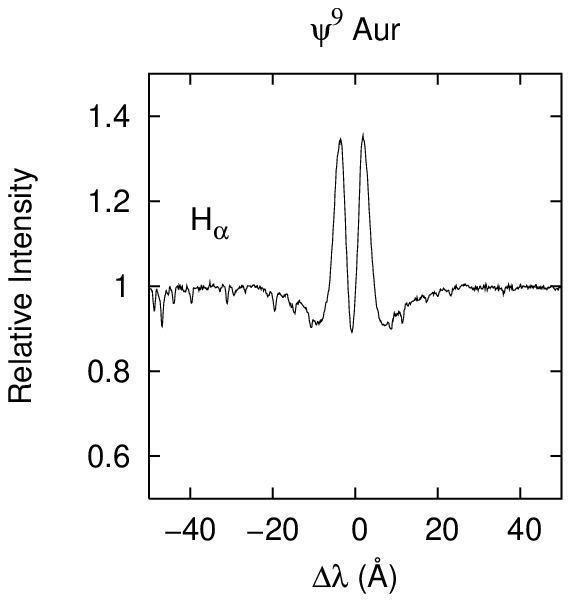}}
\resizebox{6cm}{!}{\includegraphics{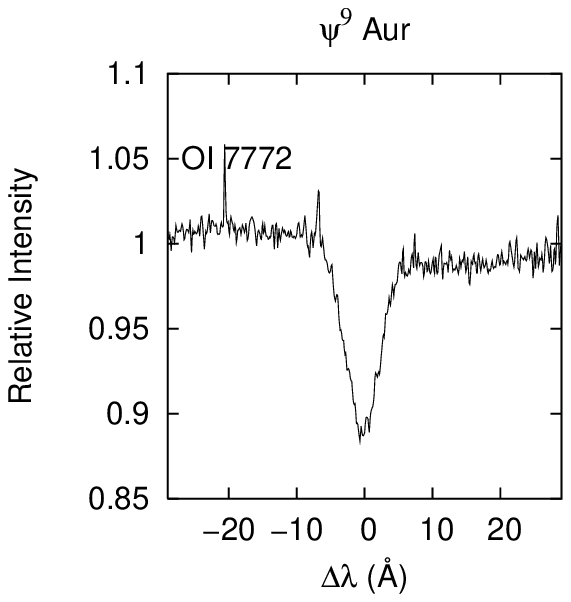}}
\caption{Profiles of {\Halpha} and {\oxir} lines of
\object{$\psi^9$~Aur}.}
\label{psiaur}
\end{figure*}

\begin{figure*}[ht]
\resizebox{6cm}{!}{\includegraphics{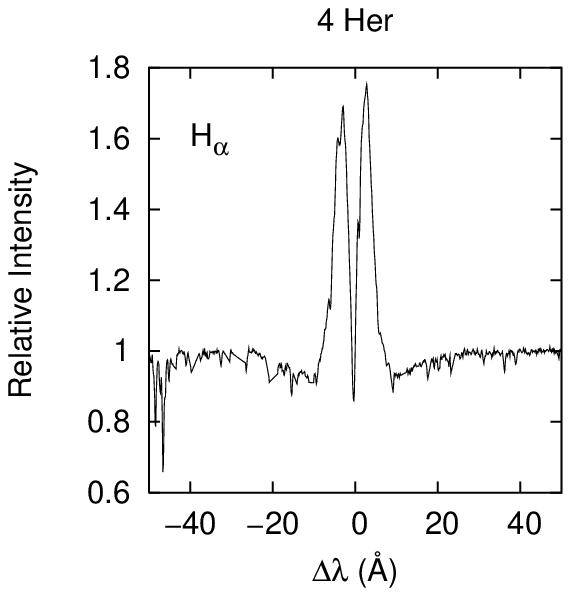}}
\resizebox{6cm}{!}{\includegraphics{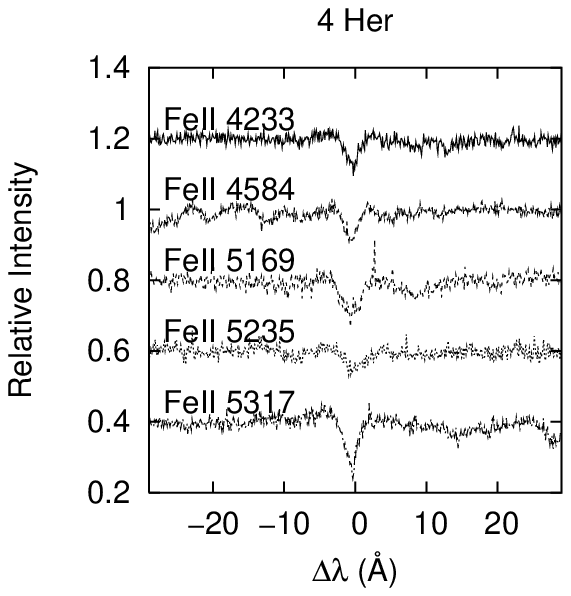}}
\resizebox{6cm}{!}{\includegraphics{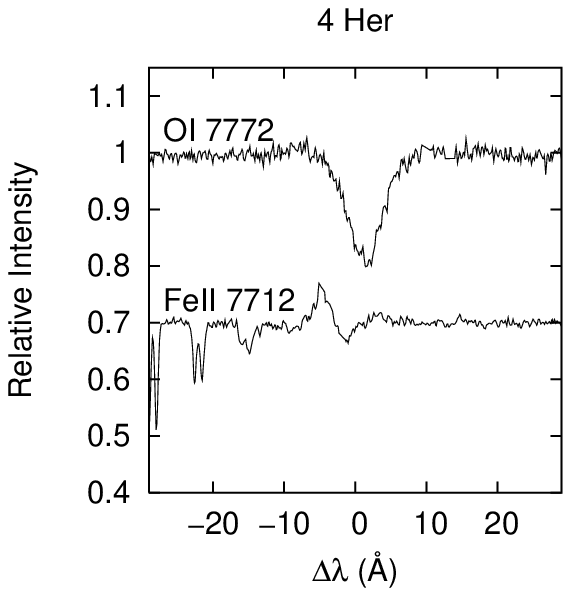}}
\caption{Profiles of {\Halpha}, {\feii} 4233, 4584, 5169, 5235, 5317,
7712\,{\AA},
and {\oxir} lines of \object{4~Her}.}
\label{4her}
\end{figure*}

\begin{figure*}[ht]
\resizebox{6cm}{!}{\includegraphics{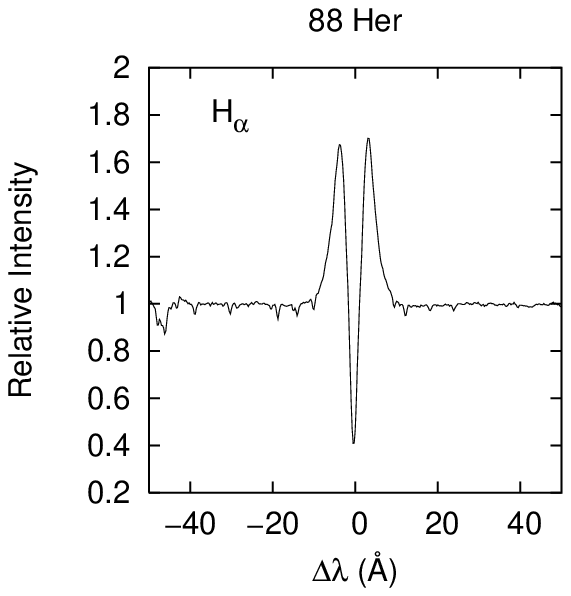}}
\resizebox{6cm}{!}{\includegraphics{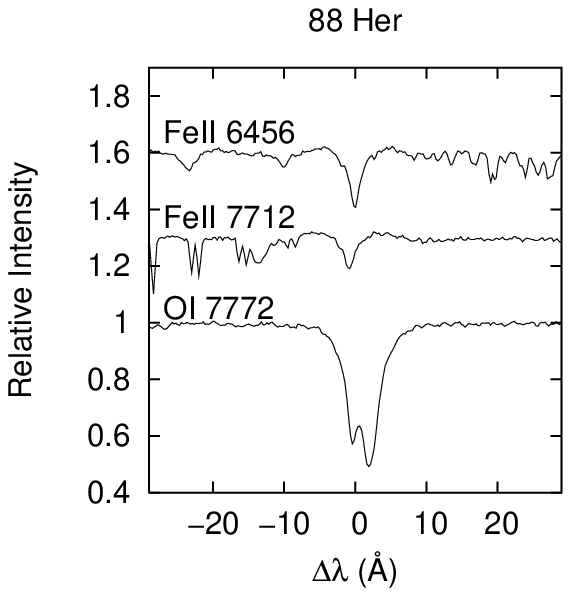}}
\caption{Profiles of {\Halpha}, {\feii} 6456, 7712\,{\AA},
and {\oxir} lines of \object{88~Her}.}
\label{88her}
\end{figure*}

\begin{figure*}[ht]
\resizebox{6cm}{!}{\includegraphics{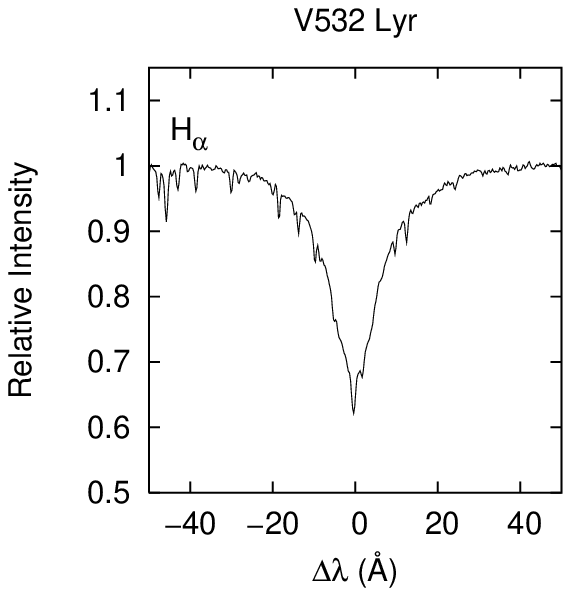}}
\resizebox{6cm}{!}{\includegraphics{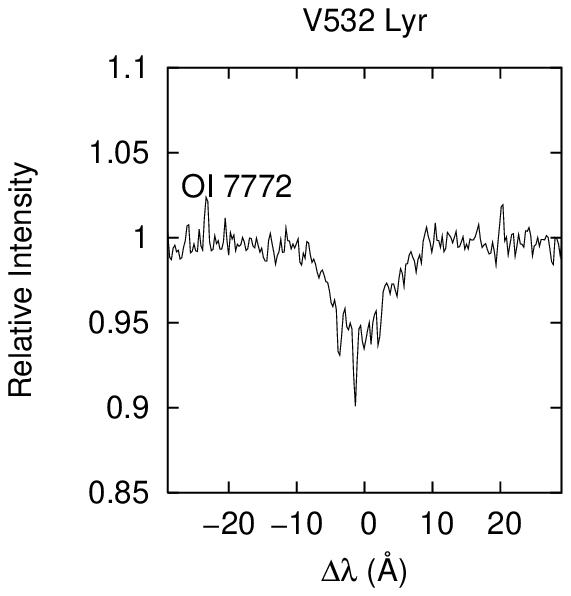}}
\caption{Profiles of {\Halpha}, {\feii} 6516\,{\AA}, and {\oxir} lines
of \object{HR~6971} (\object{HD~171406, \object{V\,532~Lyr}}).}
\label{HD171406}
\end{figure*} 

\begin{figure*}[ht]
\resizebox{6cm}{!}{\includegraphics{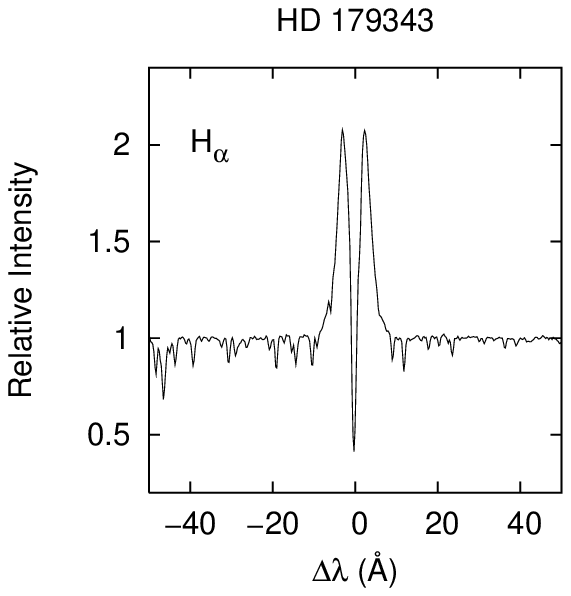}}
\resizebox{6cm}{!}{\includegraphics{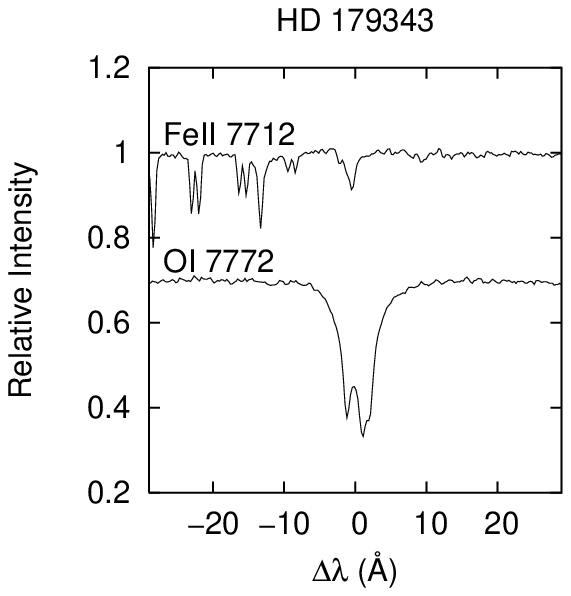}}
\caption{Profiles of {\Halpha},
{\feii} 7712\,{\AA},
and {\oxir} lines of \object{HD~179343}.}
\label{HD179343}
\end{figure*}

\begin{figure*}[ht]
\resizebox{6cm}{!}{\includegraphics{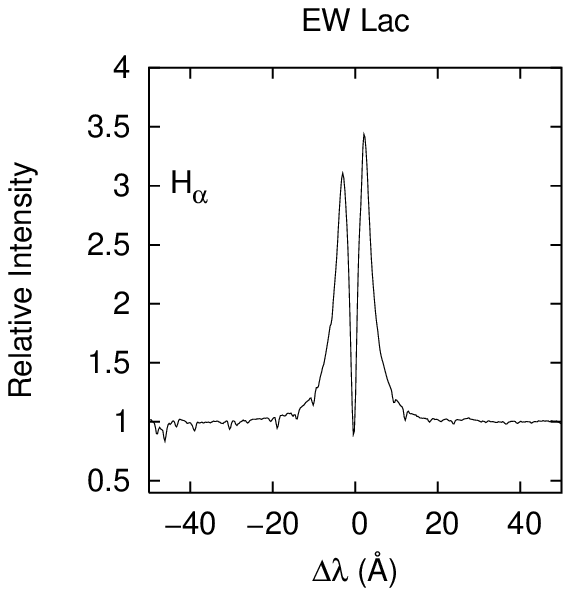}}
\resizebox{6cm}{!}{\includegraphics{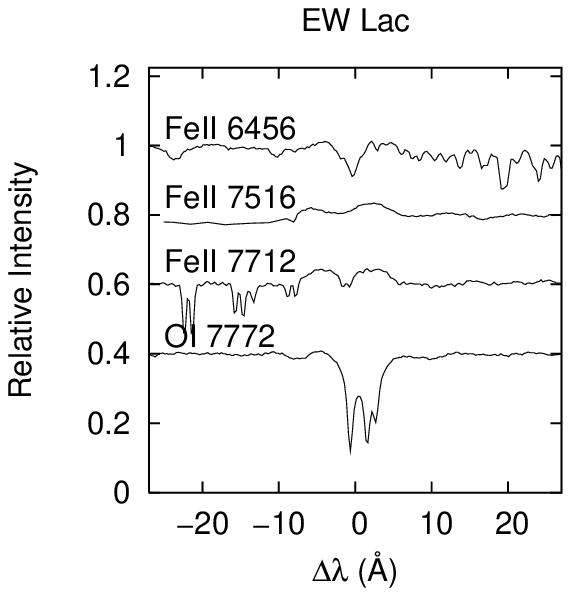}}
\caption{Profiles of {\Halpha}, {\feii} 6456,
7516, 7712\,{\AA},
and {\oxir}
lines of \object{EW~Lac}.}
\label{ewlac}
\end{figure*}

\clearpage
\subsection{{\Ab} subclass}

\begin{figure*}[ht]
\resizebox{6cm}{!}{\includegraphics{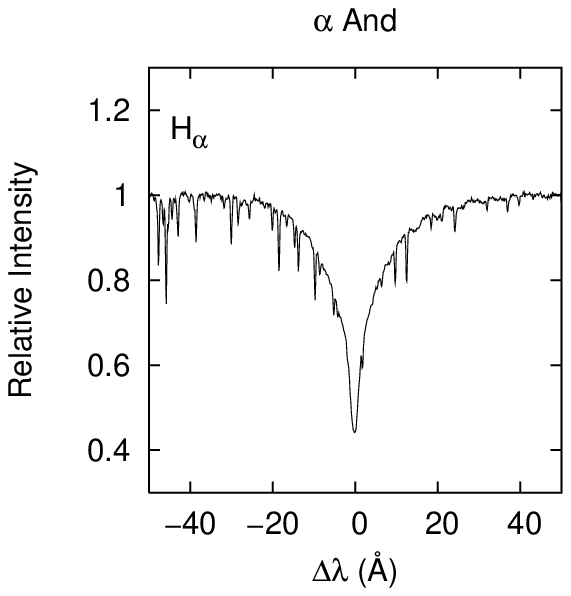}}
\resizebox{6cm}{!}{\includegraphics{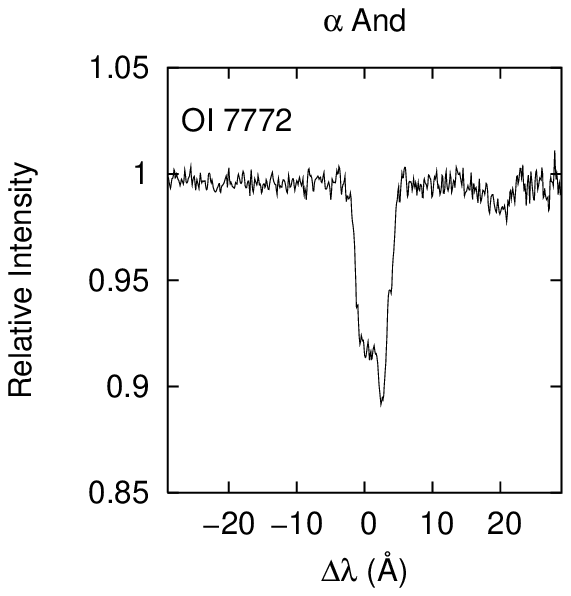}}
\caption{Profiles of {\Halpha} and {\oxir} lines of
\object{$\alpha$~And}.}
\label{alpand}
\end{figure*}

\begin{figure*}[ht]
\resizebox{6cm}{!}{\includegraphics{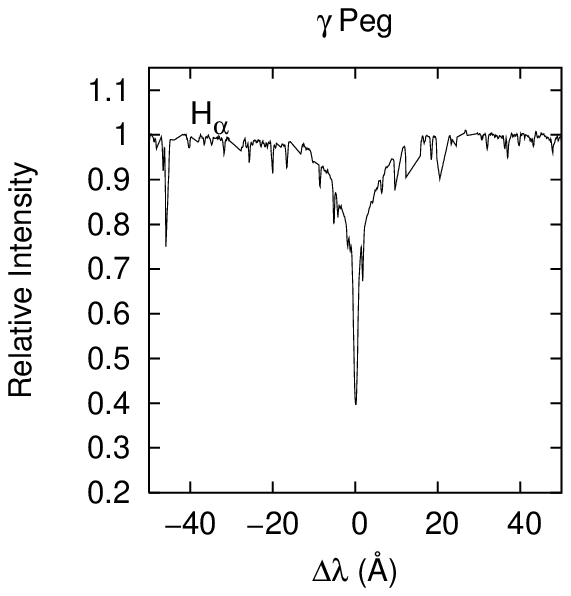}}
\resizebox{6cm}{!}{\includegraphics{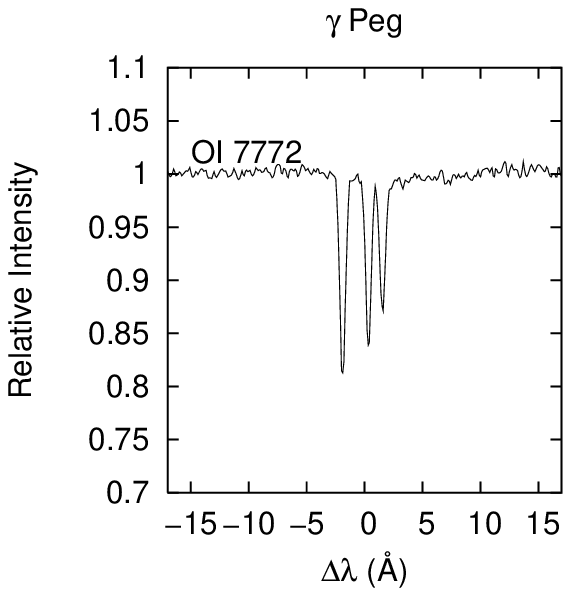}}
\caption{Profiles of {\Halpha} and {\oxir} lines of
\object{$\gamma$~Peg}.}
\label{gpeg}
\end{figure*}

\begin{figure*}[ht]
\resizebox{6cm}{!}{\includegraphics{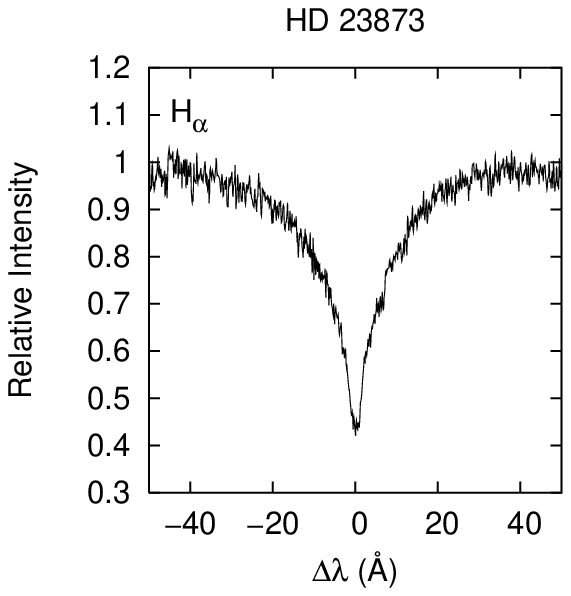}}
\resizebox{6cm}{!}{\includegraphics{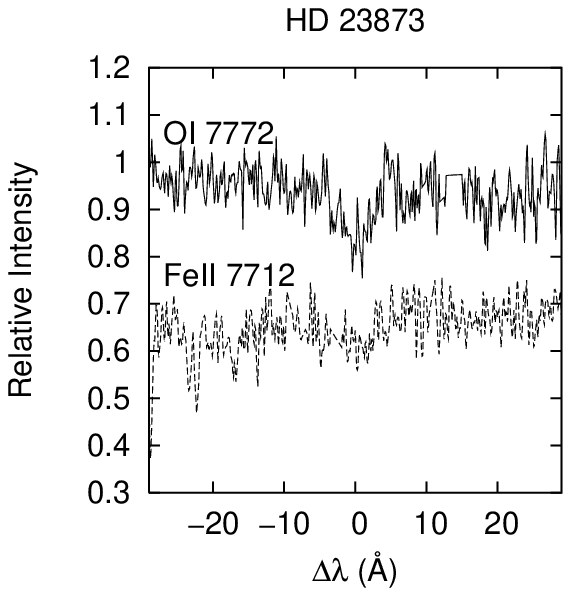}}
\caption{Profiles of {\Halpha}, {\feii} 7712\,{\AA}, and {\oxir} lines
of \object{HD~23873}.}
\label{HD23873}
\end{figure*}

\begin{figure*}[ht]
\resizebox{6cm}{!}{\includegraphics{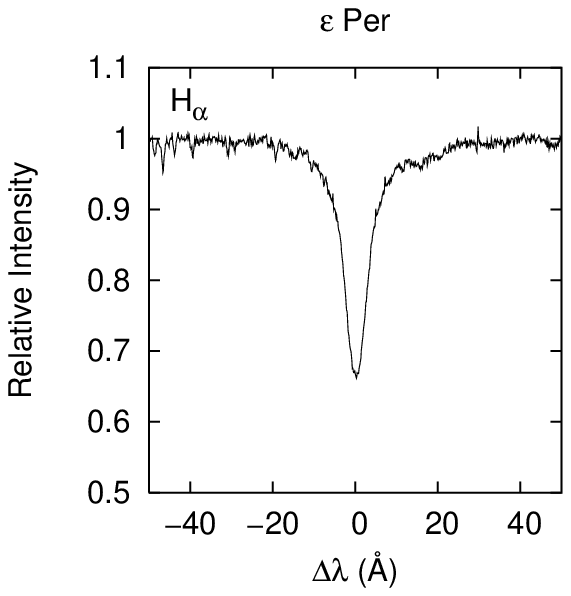}}
\resizebox{6cm}{!}{\includegraphics{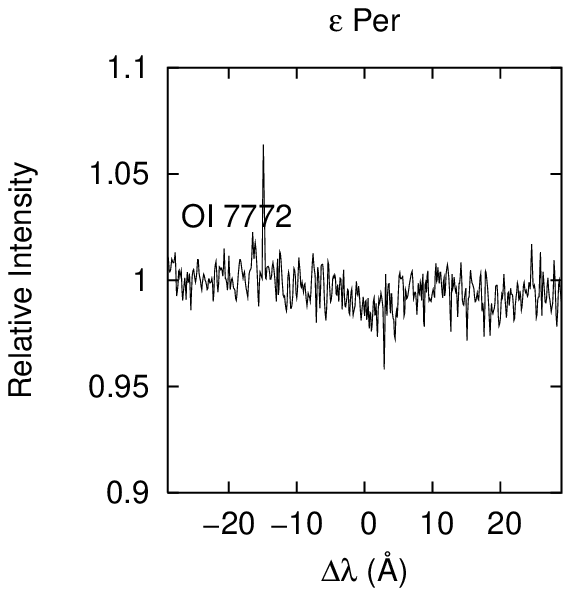}}
\caption{Profiles of {\Halpha}, {\feii} 6516\,{\AA}, and {\oxir} lines
of \object{$\varepsilon$~Per}.}
\label{epsper}
\end{figure*}

\begin{figure*}[ht]
\resizebox{6cm}{!}{\includegraphics{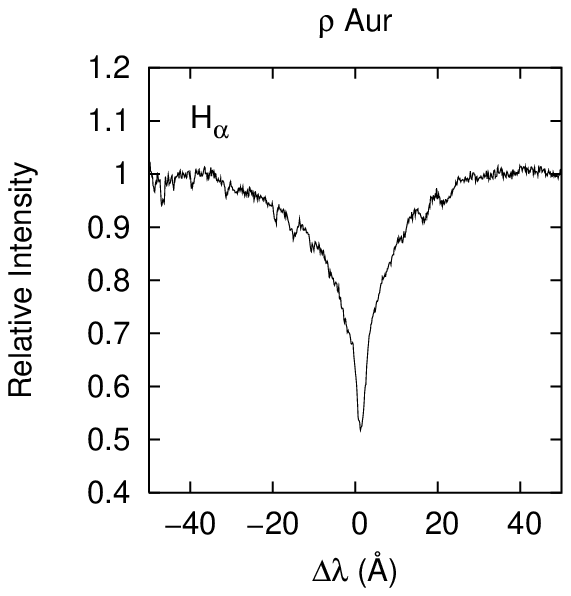}}
\resizebox{6cm}{!}{\includegraphics{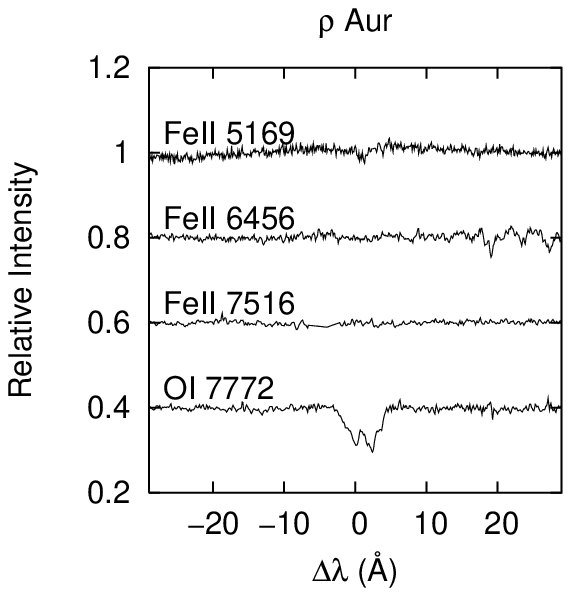}}
\caption{Profiles of {\Halpha}, {\feii} 5169, 6456, 7516,
and {\oxir} lines of \object{$\rho$~Aur}.}
\label{rhoaur}
\end{figure*}

\begin{figure*}[ht]
\resizebox{6cm}{!}{\includegraphics{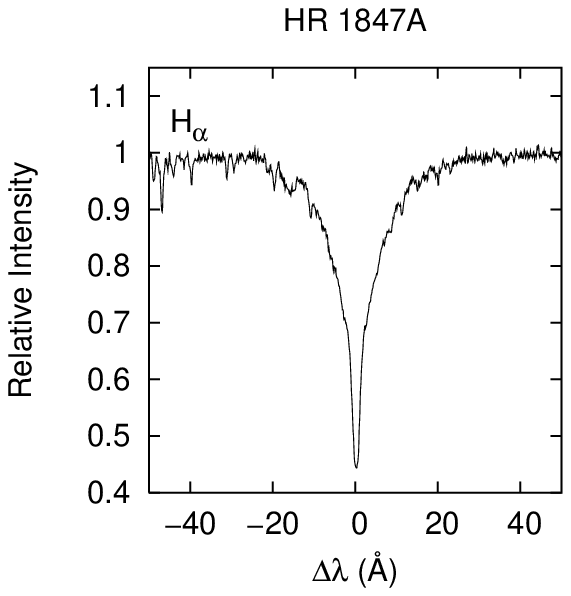}}
\resizebox{6cm}{!}{\includegraphics{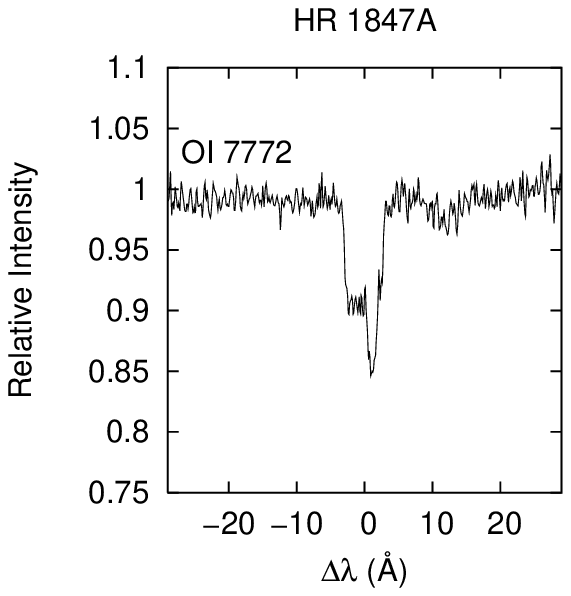}}
\caption{Profiles of {\Halpha} and {\oxir} lines of
\object{HR~1847A}.}
\label{HR1847A}
\end{figure*}

\begin{figure*}[ht]
\resizebox{6cm}{!}{\includegraphics{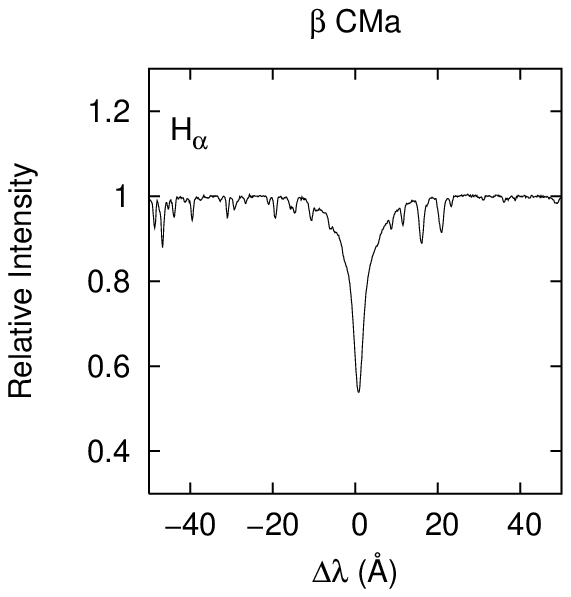}}
\resizebox{6cm}{!}{\includegraphics{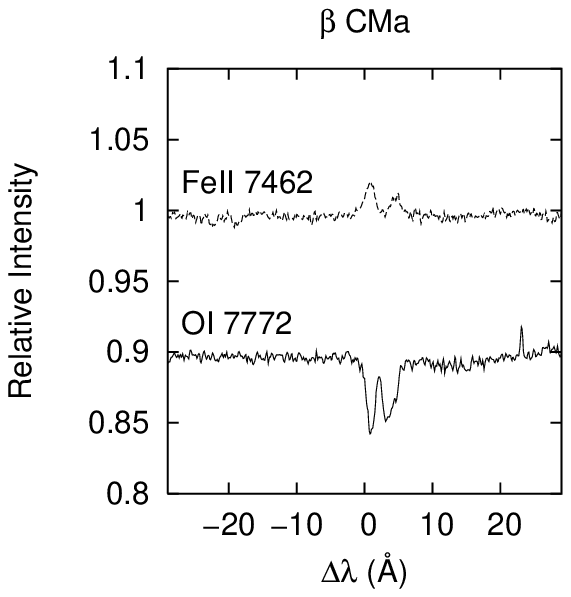}}
\caption{Profiles of {\Halpha}, {\feii} 7462\,{\AA}, and {\oxir} lines
of \object{$\beta$~CMa}.}
\label{betcma}
\end{figure*}

\begin{figure*}[ht]
\resizebox{6cm}{!}{\includegraphics{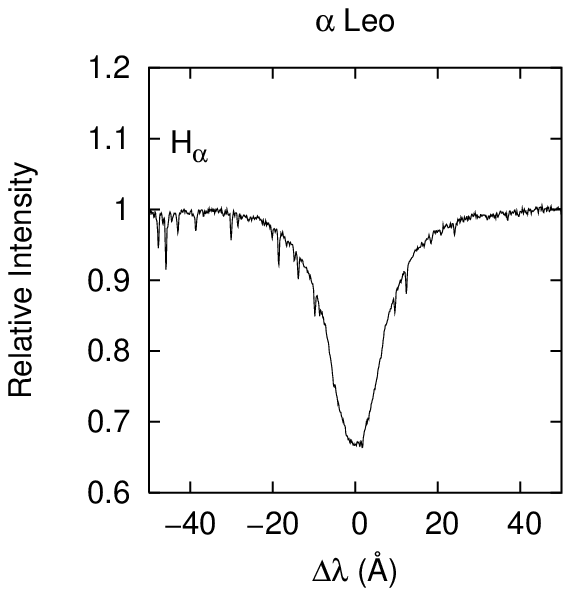}}
\resizebox{6cm}{!}{\includegraphics{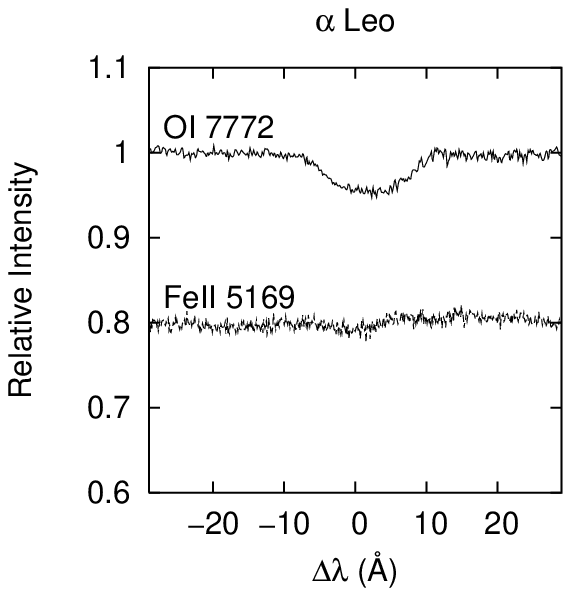}}
\caption{Profiles of {\Halpha},
{\feii} 5169\,{\AA},
and {\oxir} lines of
\object{$\alpha$~Leo}.}
\label{alpleo}
\end{figure*}

\begin{figure*}[ht]
\resizebox{6cm}{!}{\includegraphics{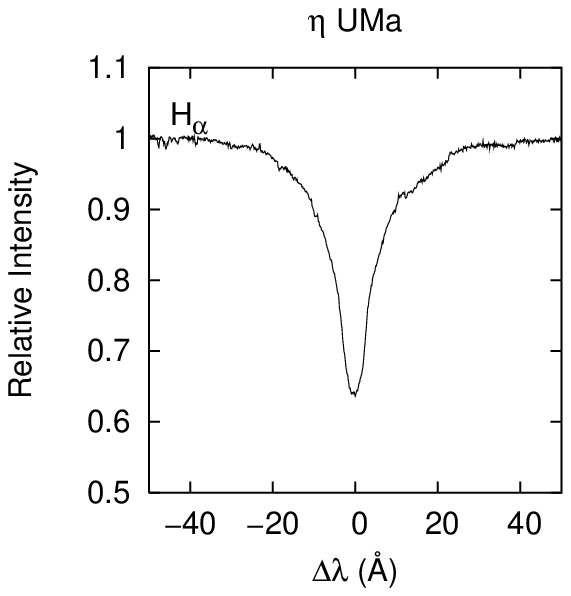}}
\resizebox{6cm}{!}{\includegraphics{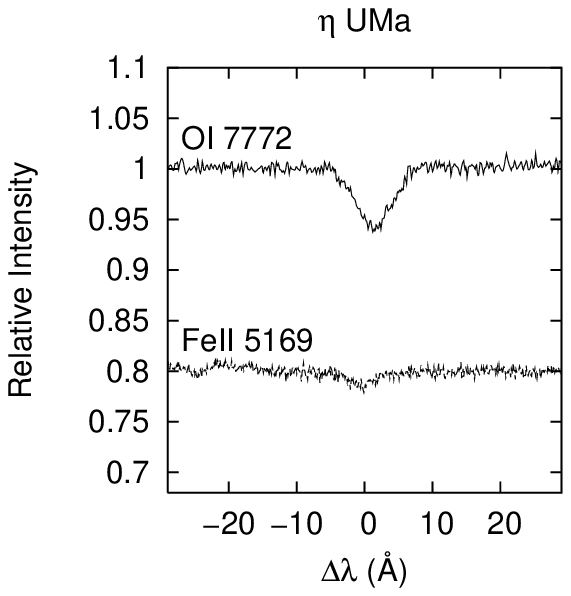}}
\caption{Profiles of {\Halpha},
{\feii} 5169\,{\AA},
and {\oxir} lines of
\object{$\eta$~UMa}.}
\label{etauma}
\end{figure*}

\begin{figure*}[ht]
\resizebox{6cm}{!}{\includegraphics{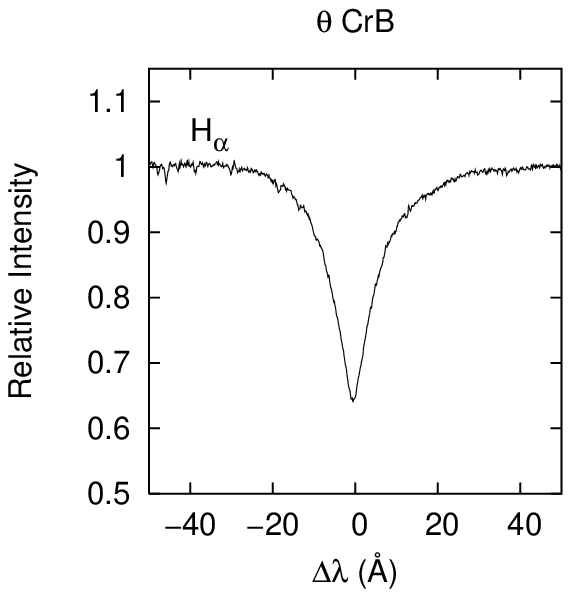}}
\resizebox{6cm}{!}{\includegraphics{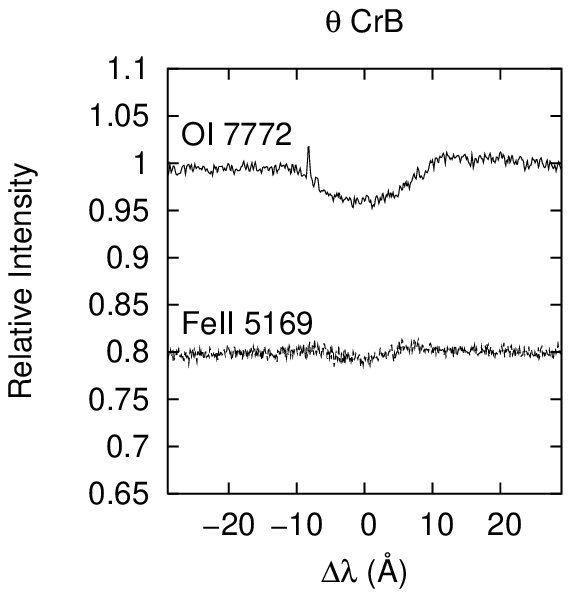}}
\caption{Profiles of {\Halpha},
{\feii} 5169\,{\AA},
and {\oxir} lines of
\object{$\theta$~CrB}.}
\label{thecrb}
\end{figure*}

\begin{figure*}[ht]
\resizebox{6cm}{!}{\includegraphics{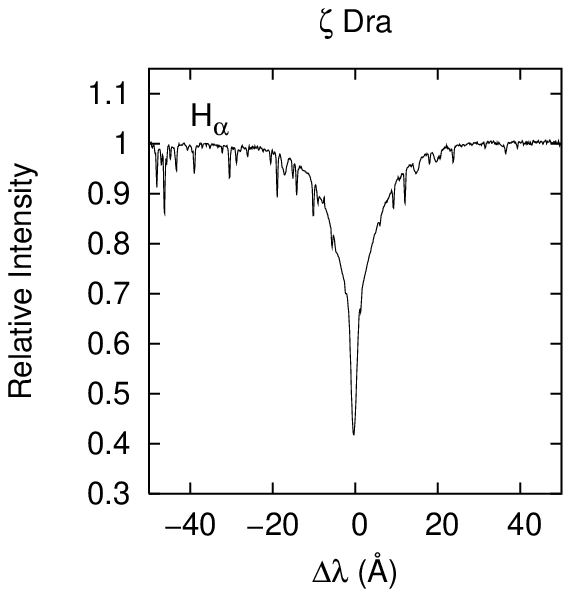}}
\resizebox{6cm}{!}{\includegraphics{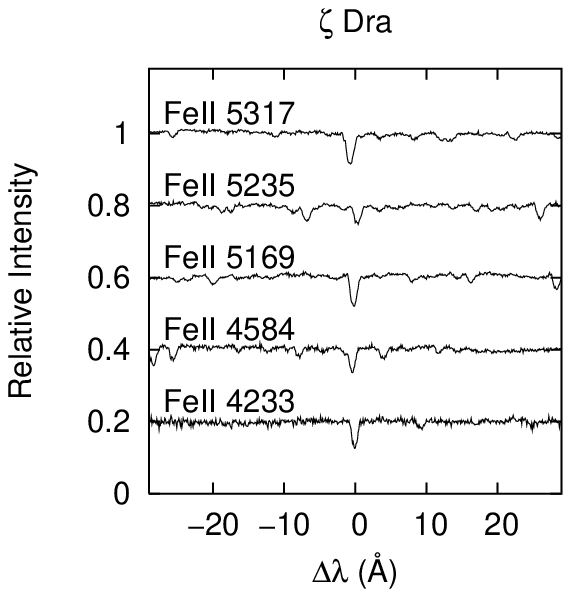}}
\resizebox{6cm}{!}{\includegraphics{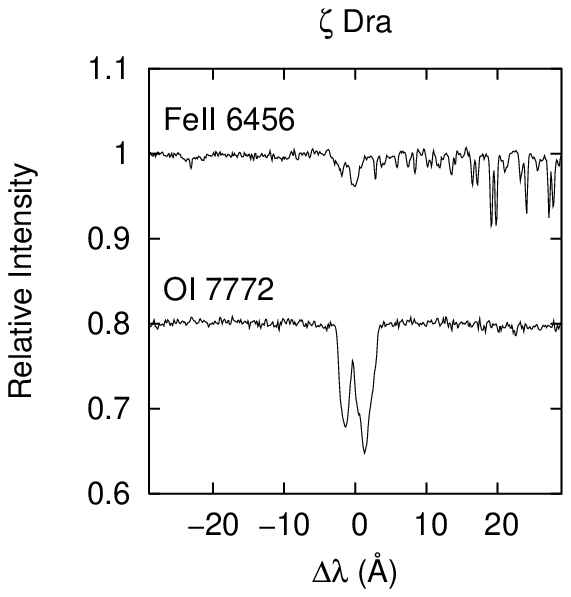}}

\caption{Profiles of {\Halpha}, {\feii} 4233. 4584, 5169, 5235, 5317, 
6456\,{\AA} and {\oxir} lines of \object{$\zeta$~Dra}.}
\label{zdra}
\end{figure*}

\begin{figure*}[ht]
\resizebox{6cm}{!}{\includegraphics{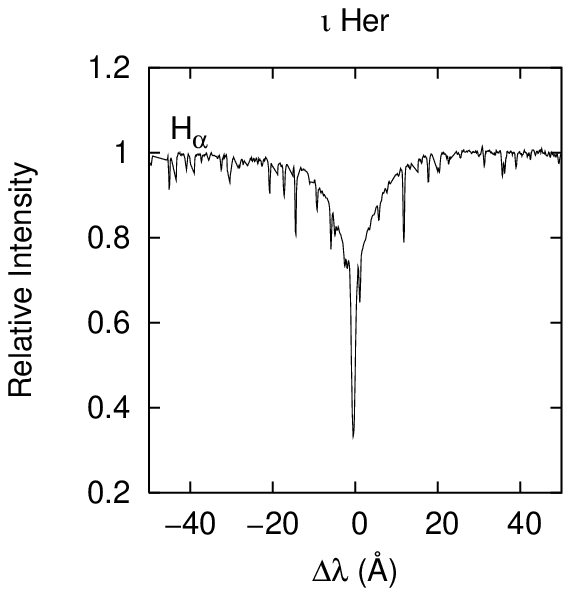}}
\resizebox{6cm}{!}{\includegraphics{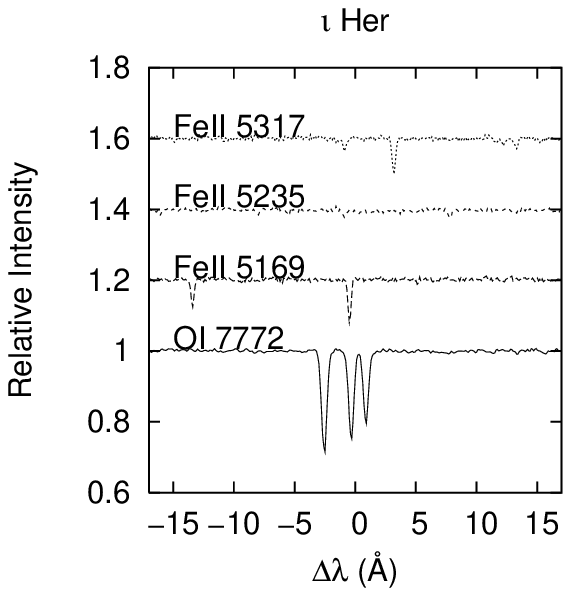}}
\caption{Profiles of {\Halpha}, {\feii} 5169, 5235, 5317\,{\AA},
and {\oxir} lines of \object{$\iota$~Her}.}
\label{iother}
\end{figure*}

\begin{figure*}[ht]
\resizebox{6cm}{!}{\includegraphics{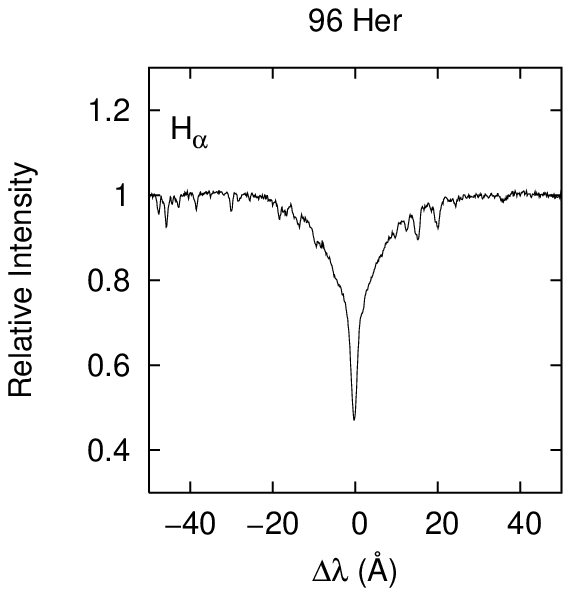}}
\resizebox{6cm}{!}{\includegraphics{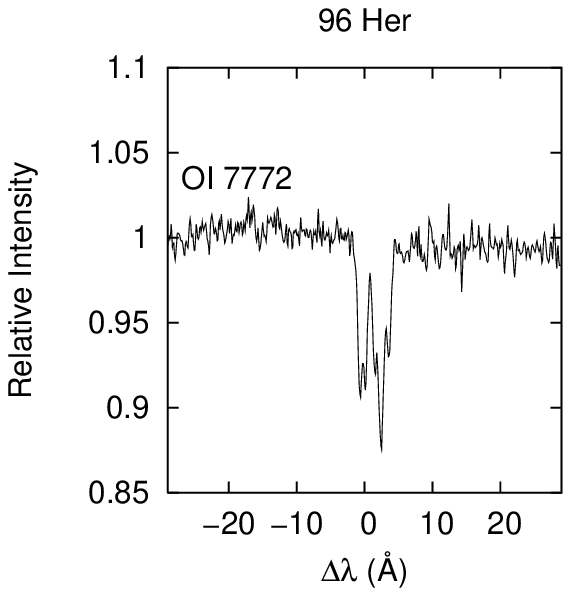}}
\caption{Profiles of {\Halpha} and {\oxir} lines of \object{96~Her}.}
\label{96her}
\end{figure*}

\end{document}

%% file: seznam04.tex
\begin{table*}
\caption{List of the B and Be stars in our sample.
References to the spectral type are denoted by superscript numbers
in parentheses, and the values of {\Teff} 
are from Theodossiou \& Danezis (\cite{ThD91}).}
\label{hvezdy}
\begin{tabular}{rrlllcccl} \hline
&&&&&&&&\\
\textbf{HD}&
\textbf{HR}&
\textbf{Name}&
\textbf{Sp. Type}&
\textbf{T$_\mathbf{eff}$}&
\textbf{JD--24 00 000}&
\textbf{JD--24 00 000}&
\textbf{Shape}\\
&&&&&&&& \\ \hline
 22780&1113 &              &B6V$^{(1)}$          &$12915\pm393 $&53217.5656&53217.5434&{\Ae}\\
 23302&1142 &17~Tau        &B6III$^{(1)}$        &$13810\pm383 $&52619.9573&      &{\Ae}\\
36408B&1847B&              &B7IV$^{(4)}$        &$12940\pm1230$&52721.8196&      &{\Ae}\\
217675&8762 &o~And         &B6III$^{(1)}$        &$13810\pm383 $&51779.0595&      &{\Ae}\\
\\
   144&   7 &10~Cas        &B9III$^{(1)}$        &$10700\pm385 $&52931.3284&53204.5171&{\Ea}\\
  6811& 335 &$\phi$~And    &B7III$^{(1)}$        &$12940\pm1230$&52566.8285&      &{\Ea}\\
 58050&2817 &OT~Gem       &B2V$^{(1)}$           &$22400\pm1393$&52697.9049&      &{\Ea}\\
 58715&2845 &$\beta$~CMi   &B8Vn$^{(1)}$         &$12120\pm623 $&52620.0981&       &{\Ea}\\
164447&6720 &V974~Her      &B8Vn$^{(1)}$         &$12120\pm623 $&53215.4018&53215.4485&{\Ea} \\
171780&6984 &              &B6V$^{(1)}$          &$15310\pm750$ &53070.6629&53182.4968&{\Ea}\\   
200310&8053 &60~Cyg        &B1V$^{(1)}$          &$25570\pm3652$&52567.9192&      &{\Ea}\\
205021&8238 &$\beta$~Cep   &B2III$^{(5)}$   &$22160\pm1145$&52113.0762&      &{\Ea}\\
216200&8690 &V360~Lac      &B3III$^{(2)}$        &$18445\pm1426$&52527.0736&      &{\Ea}\\
\\
  4180& 193 &o~Cas         &B5III$^{(1)}$        &$15310\pm750 $&52694.8389&      &{\Em}\\
  5394& 264 &$\gamma$~Cas  &B0.5IV$^{(1)}$       &$30025\pm2160$&52682.7532&      &{\Em}\\
 10516& 496 &$\phi$~Per    &B1.5(V:)e-sh$^{(2)}$ &$25570\pm3652$&52465.0435&      &{\Em}\\
 18552& 894 &              &B7IVe$^{(2)}$        &$12120\pm623 $&51954.7961&      &{\Em}\\
 22192& 1087&$\psi$~Per    &B3IIIe-sh$^{(2)}$    &$15310\pm750$ &53216.5315&53216.4905 &{\Em}\\

 23630&1165 &$\eta$~Tau    &B7IIIe$^{(2)}$       &$12940\pm1230$&52648.8068&      &{\Em}\\
 29866&1500 &              &B7IV$^{(1)}$         &$12120\pm623 $&52526.1050&      &{\Em}\\
109387&4787 &$\kappa$~Dra  &B5IIIe$^{(2)}$       &$15310\pm750 $&52720.9752&      &{\Em}\\
175863&     &BD+59 1929    &B4Ve $^{(3)}$        &$17100\pm386 $&53182.4428&53182.4642&{\Em}\\
193911&7789 &25~Vul        &B6IVe$^{(2)}$        &$14340\pm570 $&52525.9284&      &{\Em}\\
200120&8047 &59~Cyg        &B1Ve$^{(2)}$         &$25340\pm2164$&51795.9049&      &{\Em}\\
203467&8171 &6~Cep         &B2.5Ve$^{(2)}$       &$18445\pm1426$&52721.0674&      &{\Em}\\
206773&     &BD+57\degr2374&B0V:pe$^{(5)}$       &$29230\pm208 $&52484.0447&      &{\Em}\\
217891&8773 &$\beta$~Psc   &B5Ve$^{(2)}$         &$15310\pm750 $&53216.5717&53216.5525&{\Em}\\
      &     & EE~Cep       &B5:ne$\beta^{(13)}$      &              &53217.5020&53217.4567&{\Em}\\  

\\
 23862&1180 &28~Tau        &B8(V:)e-sh$^{(2)}$   &$12120\pm623 $&51797.0884&      &{\Sh}\\
 37202&1910 &$\zeta$~Tau   &B1IVe-sh$^{(2)}$     &$25570\pm3652$&51897.9234&      &{\Sh}\\
 50658&2568 &$\psi^{9}$~Aur&B6IV$^{(2)}$         &$12300\pm945 $&52721.9337&      &{\Sh}\\
142926&5938 &4~Her         &B7IVe-sh$^{(2)}$     &$12300\pm945 $&52488.8992&      &{\Sh}\\
162732&6664 &88~Her        &B7Vn$^{(1)}$         &$12915\pm393 $&53217.4040&53217.4229&{\Sh}\\
171406&6971 &V532~Lyr      &B5V$^{(1)}$          &$17100\pm386 $&53182.3954&53182.4168&{\Sh}\\
179343&     &BD+02\degr3815&B9V$^{(1)}$          &$10580\pm373 $&53216.4505&53216.4034&{\Sh}\\
217050&8731 &EW~Lac        &B3IVe-sh$^{(2)}$     &$18445\pm1426$&53217.3848&53217.3731&{\Sh}\\
\\
   358&  15 &$\alpha$~And  &B9p$^{(11)}$      &$10700\pm385 $&51779.1187&      &{\Ab}\\
   886&  39 &$\gamma$~Peg  &B2IV$^{(3)}$         &$22400\pm1393$&51771.0511&      &{\Ab}\\
 23873&     &BD+23\degr561 &B9.5V$^{(12)}$        &$10340\pm465 $&52651.0012&      &{\Ab}\\
 24760&1220 &$\epsilon$~Per&B0.5V$^{(5)}$     &$30000       $&52648.7495&      &{\Ab}\\
 34759&1749 &$\rho$~Aur    &B3V$^{(7)}$          &$15310\pm750 $&52351.9306&      &{\Ab}\\
 36408&1847A&              &B7III$^{(4)}$        &$12940\pm1230$&52720.8740&      &{\Ab}\\
 44743&2294 &$\beta$~CMa   &B1II/III$^{(5)}$     &$26105\pm1779$&52720.7954&      &{\Ab}\\
 87901&3982 &$\alpha$~Leo  &B7Vn$^{(6)}$         &$12120\pm623 $&51925.0219&      &{\Ab}\\
120315&5191 &$\eta$~UMa    &B3V$^{(7)}$          &$18445\pm1426$&52618.2262&      &{\Ab}\\
138749&5778 &$\theta$~CrB  &B6Vn$^{(1)}$         &$14340\pm570 $&52693.1422&      &{\Ab}\\
155763&6396 &$\zeta$~Dra   &B7III$^{(8)}$        &$13810\pm383 $&52002.1433&      &{\Ab}\\
160762&6588 &$\iota$~Her   &B3V SB$^{(9)}$       &$18445\pm1426$&51787.9064&      &{\Ab}\\
164852&6738 &96~Her        &B3IV$^{(10)}$        &$18445\pm1426$&52722.1490&      &{\Ab}\\
\\ \hline
\end{tabular}

{\sl References:}
(1)  -- Jaschek et al.    (\cite{Jas80});
(2)  -- Slettebak         (\cite{Sle82});
(3)  -- SIMBAD;
(4)  -- Levato            (\cite{Lev75});
(5)  -- Morgan et al.     (\cite{Mor55});
(6)  -- Murphy            (\cite{Mur69});
(7)  -- Morgan \& Keenan  (\cite{MK73});
(8)  -- Molnar            (\cite{Mol72});
(9)  -- Johnson \& Morgan (\cite{JM53});
(10) -- Lesh             (\cite{Les68});
(11) -- Cowley et al.    (\cite{Cow69});
(12) -- Mendoza          (\cite{Men56});
(13) -- Herbig			(\cite{herbig});

\end{table*}

%% file: carym.tex
\begin{table}[h]
\caption{List presented {\feii} line profiles.
For all stars the {\Halpha} and {\oxir} lines are presented as well.}
\label{cary}
\begin{tabular}{|r|c|c|c|c|c|c|c|c|c|c|c|}
\hline
\textbf{Star}&
\rotatebox{90}{\textbf{4233}}&
\rotatebox{90}{\textbf{4584}}&
\rotatebox{90}{\textbf{5169}}&
\rotatebox{90}{\textbf{5235}}&
\rotatebox{90}{\textbf{5317}}&
\rotatebox{90}{\textbf{5363}}&
\rotatebox{90}{\textbf{6432}}&
\rotatebox{90}{\textbf{6456}}&
\rotatebox{90}{\textbf{7462}}&
\rotatebox{90}{\textbf{7516}}&
\rotatebox{90}{\textbf{7712}}
\\ \hline
 HR 1847B       & & &+&+&+& & & & & & \\
 o~And          & & &+&+&+& & & & & & \\
 \hline
 10~Cas         & & & & & & & &+& & & \\
 $\phi$~And     &+&+& & & & & & & & & \\
 $\beta$~CMi    & & &+&+&+& & &+& & &+\\
 V974~Her       & & & & & & & & & & &+\\
 \hline
 o~Cas          &+&+&+&+&+& & & & &+&+\\
 $\gamma$~Cas   &+&+&+&+&+& & & & &+&+\\
 $\phi$~Per     &+&+&+&+&+& & & & &+&+\\
 HR 894         & & &+&+&+& & & & & & \\
 $\psi$~Per     & & & & & & &+&+& & &+\\
 HR 1500        & & &+&+&+& & & & & & \\
 $\kappa$~Dra   &+&+&+&+&+&+& &+& &+&+\\
 HD 175863      & & & & & & & &+& & &+\\
 25~Vul         & & &+&+&+& & & & & & \\
 59~Cyg         & & & & & & & & & &+&+\\
 6~Cep          &+&+&+&+&+& & &+& &+&+\\
 HD 206773      & & & & & & & & & &+&+\\
 $\beta$~Psc    & & & & & & & & & &+&+\\
 EE~Cep         & & & & & & & & & &+&+\\
 \hline
 28~Tau         &+&+&+&+&+& & & & &+&+\\
 $\zeta$~Tau    &+&+&+&+&+& & & & &+&+\\
 4~Her          &+&+&+&+&+& & & & & &+\\
 88~Her         & & & & & & & &+& & &+\\
 HD 179343      & & & & & & & & & & &+\\
 EW~Lac         & & & & & & & &+& &+&+\\
 \hline
 HD 23873       & & & & & & & & & & &+\\
 $\rho$~Aur     & & &+& & & & &+& &+& \\
 $\beta$~CMa    & & & & & & & & &+& & \\
 $\alpha$~Leo   & & &+& & & & & & & & \\
 $\eta$~UMa     & & &+& & & & & & & & \\
 $\theta$~CrB   & & &+& & & & & & & & \\
 $\zeta$~Dra    &+&+&+&+&+& & &+& & & \\
 $\iota$~Her    & & &+&+&+& & & & & & \\
\hline
\end{tabular}
\end{table}